\documentclass[aps,prl,superscriptaddress,reprint,floatfix]{revtex4-2}
\usepackage{graphicx}
\usepackage{amsmath}
\usepackage{amssymb}
\usepackage{mathrsfs}
\usepackage[T1]{fontenc}
\usepackage{calligra}
\usepackage{array}
\usepackage{xcolor}
\usepackage{float}
\usepackage{soul}
\usepackage{bm}
\usepackage[toc,page]{appendix}
\usepackage{braket}
\usepackage{wrapfig}
\usepackage[colorlinks=true, linkcolor=blue, urlcolor=blue,citecolor=blue]{hyperref}

\hypersetup{colorlinks,linkcolor=blue,citecolor=blue,urlcolor=blue}

\usepackage{babel}
\begin{document}
\title{Extrinsic Orbital Hall Effect: Orbital Skew Scattering and Crossover Between Diffusive and Intrinsic Orbital Transport}
\author{Alessandro Veneri}
\affiliation{School of Physics, Engineering and Technology and York Centre for Quantum Technologies, University of York, York YO10 5DD, United Kingdom}
\author{Tatiana G. Rappoport}
\affiliation{Physics Center of Minho and Porto Universities (CF-UM-UP),Campus of Gualtar, 4710-057, Braga, Portugal}
\affiliation{International Iberian Nanotechnology Laboratory (INL),
Av. Mestre José Veiga, 4715-330 Braga, Portugal}
\author{Aires Ferreira}
\email{aires.ferreira@york.ac.uk}
\affiliation{School of Physics, Engineering and Technology and York Centre for Quantum Technologies, University of York, York YO10 5DD, United Kingdom}

\begin{abstract}

Despite the recent success of identifying experimental signatures of the orbital Hall effect (OHE), the research on the microscopic mechanisms behind this unique phenomenon is still in its infancy. Here, using a gapped 2D Dirac material as a model system of the OHE, we develop a microscopic theory of orbital transport which captures extrinsic disorder effects non-perturbatively. We show that it predicts several hitherto unknown effects, including (i) a strong dependence of the orbital Hall conductivity with the strength and symmetry of the impurity scattering potential, and (ii) a smooth crossover from intrinsic to extrinsic OHE as a function of the Fermi energy and impurity density. In contrast to previous (perturbative) studies, the OHE is found to exhibit bona fide diffusive behavior in the dilute impurity limit, which we trace back to the dominance of skew scattering-type processes. More generally, we argue that the newly unveiled orbital skew scattering mechanism governs the diffusive OHEs of a large class of 2D materials even when the crystal structure is inversion-symmetric. Our work unveils the crucial nature of non-perturbative vertex corrections for a complete description of orbital transport and confirms common short-range impurities as key enablers of the OHE.

\end{abstract}
 
\maketitle

The transport of orbital angular momentum (OAM) in solids has garnered significant interest for its fundamental role in understanding quantum dynamics in spin-orbit coupled systems and its potential for device applications \cite{Go-Review, Ding2020, Sala2022,Jo2024}. In materials with weak spin-orbit coupling, charge-neutral orbital currents can be generated electrically via the orbital Hall effect (OHE) \cite{Kontani2008,Go2018,Salemi2022}, first proposed in 2005 \cite{OHEBernevig} and recently observed in experiments on light metals \cite{Choi2023,Lyalin2023,Sala2023}. This development is steering spintronics in new directions, with studies addressing orbital torques, ultra-fast OAM transport, and more \cite{go2020,Lee2021,Ding2024,Han2022,Busch2024,Seifert2023,Kumar2023,Xu2024,Go2023,Manchon2023,Han2023b,Santos2023,Hayashi2024,Johansson2021,ElHamdi2023}.

Efforts to unravel the microscopic mechanisms of the OHE have primarily focused on intrinsic transport driven by momentum-space orbital textures, often linked to quantum geometric effects \cite{Essay_QG_PhysRevLett.131.240001}. However, theoretical estimates based solely on intrinsic mechanisms differ significantly from experimental results in titanium thin films, suggesting that disorder plays a critical role in the relaxation of nonequilibrium orbital densities \cite{Choi2023}. Recent theoretical work supports this view, showing that thermal disorder at room temperature is an important piece of the OHE puzzle \cite{Belashchenko_23,Rang_24,Son_2024}. A pressing challenge moving forward is to understand how disorder, particularly short-range defects and impurities, affect the generation of OAM currents. Short-range defects are ubiquitous in metals and other OHE candidate materials \cite{Sala2023} and may enable efficient mechanisms of extrinsic orbital transport. A strong contender, hitherto unconsidered, is orbital skew scattering (i.e. the orbital analog of Mott scattering), whereby an applied electric field causes wave packets with opposite OAM (e.g., $L_z=\pm \hbar$) to scatter asymmetrically, resulting in transverse OAM flow. Two such mechanisms may contribute to the OHE without the need for spin-orbit coupling: (i) asymmetric scattering due to impurities with non-trivial orbital-space structure, and (ii) asymmetric scattering enabled by the orbital texture of wavefunctions. The former may induce OHE in otherwise orbital-inactive systems, while the latter is a band-driven mechanism whose spin analog has been found to emerge in 2D materials with broken spatial symmetries \cite{Offidani2017,Perkins_2024}.

Despite the increasing interest surrounding the OHE, only a few studies have systematically examined the role of disorder \cite{Pezo2023,liu_dominance_2023, Barbosa2023, Tang2024, Canonico2024}. Although a coherent picture has not yet emerged, this early work suggests that a non-perturbative treatment of disorder at some level of theory will be crucial to unlock the extrinsic OHE. Our purpose in this Letter is to fill this gap. The key issue concerns the exact nature of the vertex corrections to the orbital Hall conductivity (OHC). Specifically, Ref. \cite{liu_dominance_2023} predicts an extrinsic OHE insensitive to the disorder parameters, while Ref. \cite{Tang2024} finds that vertex corrections of the OHC vanish entirely for short-range impurities. In contrast, numerical real-space calculations carried out for small system sizes clearly show a disorder-dependent OHE \cite{Canonico2024}. Although real space numerics can simulate tight-binding models of arbitrary complexity \cite{kite}, reaching the interesting diffusive regime of macroscopic systems at low impurity densities remains a considerable challenge \cite{Castro_24}. Here, diagrammatic linear-response techniques can prove exceedingly useful, as they capture the diffusive regime by design. Indeed, we will show below that a formulation beyond the standard perturbative approach solves the conundrum of the vertex corrections and unveils a rich OHE phenomenology, where the microscopic details of the disorder landscape occupy the center stage. 

\begin{figure}
	\begin{centering}
		\includegraphics[scale=0.47]{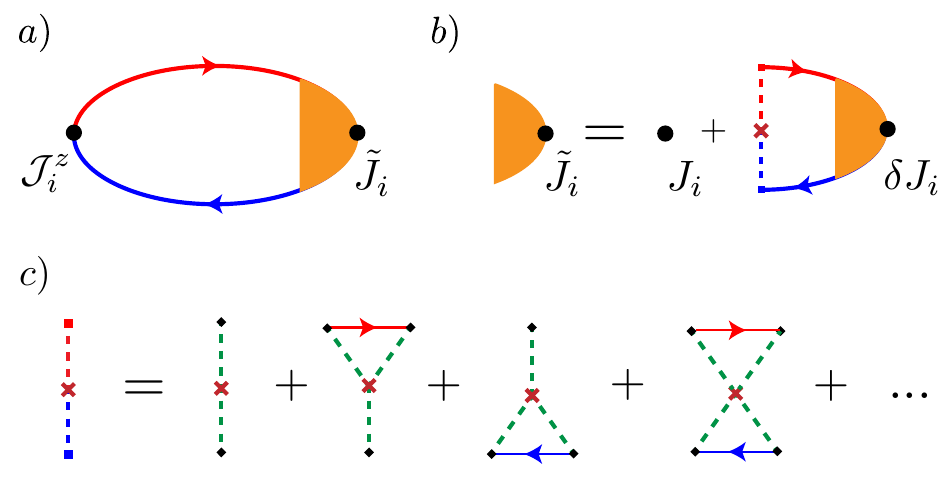}\caption{Diagrammatic technique: (a) the extrinsic OHC; (b) the disorder-renormalized charge current vertex function; and (c) the $T$-matrix expansion of this work. In the popular ladder approximation, only the first diagram in (c)   is retained. This  captures side-jump processes perturbatively, but misses out the skew scattering mechanism as well as the strong scattering regime. Solid (dashed) lines denote disorder-averaged Green’s function (single-impurity potential insertions), while red/blue indicates advanced/retarded sectors.}\label{fig:01}
		\par\end{centering}
\end{figure}

\textit{Setting the stage}. ---We explore a system of massive 2D Dirac electrons, a prototypical model for 2D materials with broken inversion symmetry. We note, however, that the extrinsic OHE and its main driving force (see below) are expected to be universal features of a large class of orbital-active 2D materials, including centrosymmetric systems. Orbital physics in 2D materials \cite{Canonico2020,Canonico2020b,Bhowal2020,Xue2020,Cysne2021,Bhowal2021,Cysne2022,Pezo2022,Cysne2023,Busch2023} such as dichalcogenide monolayers, has previously been linked to phenomena like orbital Hall insulating phases \cite{Canonico2020b,Cysne2021} and OAM-carrying in-gap edge states \cite{Costa2023}.  To incorporate disorder into the picture, we employ a diagrammatic technique wherein electron-impurity scattering events responsible for vertex corrections to the OHC are described via a \textit{generalized ladder} series of Feynman diagrams \cite{Milletari2016}. It accomplishes the exact resummation of the infinite series of 2-particle non-crossing diagrams ("ladder", "$Y$", "$X$", etc.; see Fig. \ref{fig:01}). Hence, all the extrinsic mechanisms triggered by single-impurity scattering events (e.g., semiclassical skew scattering, asymmetric scattering precession \cite{Huang_16_ASP} and side jumps) are captured in a \textit{fully nonperturbative}  fashion \cite{Milletari2016}. Previous applications include the spin Hall effect (SHE) in twisted 2D heterostructures \cite{Perkins_2024} and the anomalous Hall effect in magnetized 2D materials \cite{Offidani_2018_AHE}. Armed with this formalism, we uncover the dominant OHE mechanisms (most notably, orbital skew scattering) and construct a phase diagram of OAM transport spanning extrinsic and intrinsic regimes. The crucial role played by the symmetry of the scattering centers will also be addressed.

\textit{Theory}.---The single-particle Hamiltonian around the $K$($K^\prime$) point in the valley-isotropic basis  reads as
\begin{equation}
H_\tau=v\,\boldsymbol{\sigma}\cdot\mathbf{p}+\tau \Delta\,\sigma_{z}+V_{\text{dis}}(\mathbf{x})\,
\label{eq:ham}
\end{equation}
where $v$ is the bare Fermi velocity of 2D massless Dirac fermions, $\mathbf{p}=-i\hbar\boldsymbol{\nabla}$ is the momentum operator,  $\boldsymbol{\sigma}$ is the vector of pseudospin Pauli matrices and $\tau=\pm 1$ is the valley index. Moreover, $\Delta$ is a staggered on-site potential leading to an energy gap $E_g=2\Delta$ and $V_{\text{dis}}(\mathbf{x})$ describes the disorder landscape. To get a broader picture of the extrinsic OHE, here we shall consider a generalized short-range impurity model. Specifically, $V_{\text{dis}}(\mathbf{x})=\sum_{i} M_{\textrm{dis}}(u_0,u_z)\,\delta(\mathbf{x}-\mathbf{x}_{i})$, where $M_{\textrm{dis}}(u_0,u_z)=u_{0}\,\sigma_{0}+u_{z} \,\tau \sigma_{z}$, $\{\mathbf{x}_{i}\}$ is the set of impurity positions ($i=1,...,N$), $u_{0(z)}$ is the strength of the scalar (staggered) component of the scattering  potential, and $\sigma_0$ is the $2\times 2$ identity matrix. Our main quantum observable of interest is the orbital-current operator   $\mathcal{J}_{i}^{z}=\frac{1}{2}\{j_i,L_{z}\}$, where $j_i=v\, \sigma_i$ (with $i=x,y$) is the particle current operator, $L_z$ is the $\hat z$ component of the OAM operator and $\{\cdot,\cdot\}$ denotes the anticommutator. $L_z$  has the following momentum-space representation in the valley-isotropic basis $L_{z} (\textbf{k})=-\hbar\,\sigma_{0} \tau \Delta m_{e}v^{2}/E_{\mathbf{k}}^{2}$   \cite{Bhowal2021}, where $m_e$ is the electron mass, $E_{\mathbf{k}}=\sqrt{\hbar^2 v^{2}k^{2}+\Delta^{2}}$ is the energy dispersion, and $k=|\textbf{k}|$ is the wavevector measured from a valley. We note that wavepackets centered at the $K$ and $K^\prime$ points carry opposite OAM due to time-reversal ($\mathcal T$) symmetry. Because the impurity Hamiltonian in our model is diagonal in valley space (i.e. intervalley scattering terms are neglected), the total OHC is two times the OHC of a single valley. In the following, we work in the $K$-valley sector ($ H \equiv H_{\tau=1}$) and introduce a valley degeneracy factor ($g_v=2$) when required. 
 
The extrinsic contribution to the linear-response OHC in the dilute impurity limit is governed by the Fermi-surface (type I) term of the Kubo-Streda formula:
\begin{equation}
\mathcal{\sigma}_{ij}^{\textrm{OHE}}(\varepsilon)=\frac{g_{v}g_{s}}{2\pi}\int d\mathbf{k}\,\textrm{tr}\,\left[\mathcal{J}_{i}^{z}(\mathbf{k})\,\left\langle G_{\varepsilon}^{+}\,J_{j}\,G_{\varepsilon}^{-}\right\rangle (\textbf{k})\right],
 \label{eq:Kubo}
\end{equation}
where  $J_{i}=-e j_{i}$  is the charge current operator ($e>0$),   $G_{\varepsilon}^{\pm}=(\varepsilon-H  \pm i 0^{+})^{-1}$
is the retarded($+$)/advanced($-$) Green's function at the Fermi energy $\varepsilon$, $g_s=2$ is the spin-degeneracy factor,  $\left\langle ...\right\rangle $ is the disorder average, $\hbar \equiv 1$, 
and the trace is taken over the pseudospin degree of freedom. The expression inside angular brackets can be cast as $\left\langle G_{\varepsilon}^{+}\,J_{j}\,G_{\varepsilon}^{-}\right\rangle \rightarrow\mathcal{G}_{\varepsilon}^{+}(\mathbf{k})\,\tilde{J}_{j}\,\mathcal{G}_{\varepsilon}^{-}(\mathbf{k})\equiv2\pi\varrho_j(\varepsilon,\mathbf{k})$, where $\mathcal{G}_{\varepsilon}^{\pm}(\mathbf{k})$ are disorder-averaged Green's functions and  $\tilde{J}_{j}$ is the disorder-renormalized charge current operator obtained by solving the Bethe-Salpeter equation in Fig. \ref{fig:01}(b). Explicitly, we have: $\mathcal{G}_{\varepsilon}^{\pm}(\mathbf{k})=(\varepsilon-H_{0}(\mathbf{k})-\Sigma_{\varepsilon}^{\pm})^{-1}$, where $H_{0}(\mathbf{k})$ is the disorder-free Hamiltonian, $\Sigma_{\varepsilon}^{\pm}=n \, T^{\pm}(\varepsilon) $ is the disorder self-energy, $T^{\pm}(\varepsilon)$ is the single-impurity $T$ matrix, and $n$ is the impurity density in the thermodynamic limit ($N\rightarrow \infty$); see Ref. \cite{Supplementary_material} for detailed expressions of $\mathcal{G}_{\varepsilon}^{\pm}(\mathbf{k})$, $T^{\pm}(\varepsilon)$ and $\tilde J_j(\varepsilon)$. All together, the knowledge of the renormalized vertex yields the extrinsic OHC via $\mathcal{\sigma}_{ij}^{\textrm{OHE}}(\varepsilon)= g_s g_v \int d\mathbf{k}\,\textrm{tr}\,\left[\mathcal J_{i}^{z}(\mathbf{k})\varrho_j(\varepsilon,\mathbf{k})\right]$. This elegant expression shows that $\varrho_j(\varepsilon,\mathbf{k})$ plays the role of a $\textbf{k}$-resolved density matrix encapsulating the linear response of the system, and therefore that the \textit{structure of the renormalized vertex is key}\,\cite{Supplementary_material}. Finally, the intrinsic OHC, $\sigma_{\textrm{OHE}}^{\textrm{int}}$, is obtained by momentum space integration of the orbital Berry curvature  \cite{Bhowal2021}. In what follows, we assume that the electric field driving the OHE is applied along the $\hat x$ axis  and define $\sigma_{\text{OHE}}^{\text{dis.}}\equiv\sigma_{yx}^{\text{OHE}}$ (due to symmetry of the model, we also have $\sigma_{xy}^{\text{OHE}}=-\sigma_{yx}^{\text{OHE}}$). 
 
\textit{Results and discussion}.---We consider two types of scattering centers: (i) conventional scalar impurities  and (ii)     impurities with $u_z\neq 0$. Both cases are realistic and commonly realized. A simple example is a top impurity with $C_{3v}$ symmetry  (e.g. an ad-atom chemisorbed on an $A$- or $B$-type site in graphene \cite{Basko_08,Bena_08,Pachoud_14}). Its localized nature around a site belonging to a particular sublattice ($A$ or $B$) implies $u_z \approx \pm  u_0$,  so that the projection of $M_{\text{dis}}$ on the opposite sublattice ($B$ or $A$) vanishes or is strongly suppressed. In contrast, hollow-position impurities enjoy 6-fold rotational symmetry and thus generate purely scalar potentials ($u_z=0$) \footnote{For group VI dichalcogenide monolayers, the picture is more involved due to the different sublattice and orbital content of the basis functions of the effective model.}. This justifies our use of a generalized disorder model, and will allow us to draw a number of conclusions regarding the nature of possible OHEs. We first focus on scalar impurities, as they are the most symmetric ones. To help uncover the main driver of the extrinsic OHE, we expand Eq. (\ref{eq:Kubo}) in powers of the $n$ [or equivalently, $1/(\varepsilon \tau_0)$, where $\tau_0 \propto n^{-1}$ is a typical (charge) scattering time]. The leading term of the expansion $\mathcal O(n^{-1})$ encodes the semiclassical response:
\begin{equation}
    \begin{split}
            \sigma_{\textrm{OHE}}^{\textrm{s.c.}} =  & \chi_{\varepsilon}\,\frac{ 2e m_{e}v^{2}}{\pi n}  \theta(|\varepsilon|-|\Delta|)\left[\frac{\Delta^{2}}{u_{0}\varepsilon^{2}}\frac{f_{1}(\varepsilon,\Delta)}{f_{2}(\varepsilon,\Delta)}\right.  +\\
            & \left.\frac{4\Delta^{4}f_{1}(\varepsilon,\Delta)\log\left(\frac{\Lambda^{2}}{\varepsilon^{2}-\Delta^{2}}-1\right)}{\pi v^{2}\varepsilon\,[f_{2}(\Delta,\varepsilon)]^{3/2}}+\mathcal O(u_0)\right],
    \end{split}
\label{eq:OHE_scalar}
\end{equation}
where $\chi_{\varepsilon}=\text{sign}(\varepsilon)$, $f_{1}(\varepsilon,\Delta)=(\epsilon^{2}-\Delta^{2})^{2}$, $f_{2}(\Delta,\varepsilon)=\left(\varepsilon^{2}+3\Delta^{2}\right)^{2}$, and $\Lambda \gg \varepsilon, \Delta $ is an energy cutoff used to regularize diverging integrals appearing at higher order (typically $\Lambda \approx 10 $ eV, but the results are little sensitive to actual choice of $\Lambda$  \cite{Ferreira2011}). The validity of the analytical $u_0$-expanded result [Eq. (\ref{eq:OHE_scalar})] (accurate for $|u_0|$ up to $\approx 0.2$ eV\,nm$^2$)  is discussed in the Supplemental Material \cite{Supplementary_material}.

\begin{figure}
	\begin{centering}
		\includegraphics[scale=0.5]{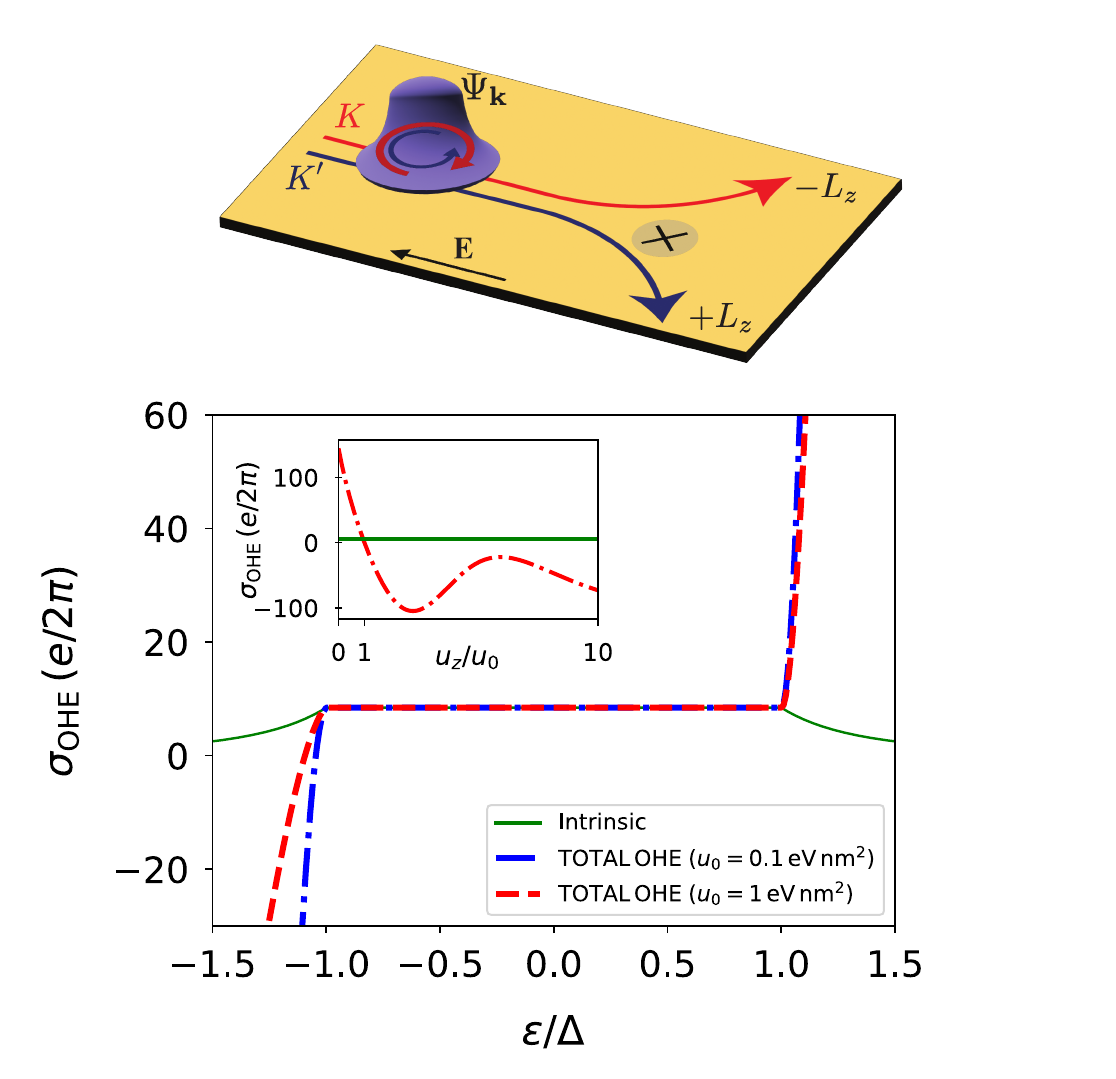}\caption{Top panel: Illustration of the orbital skew scattering mechanism leading to a transverse net flow of OAM. Bottom panel: Fermi energy dependence of the total OHC ($\sigma_{\textrm{OHE}}^{\textrm{tot}}=\sigma_{\textrm{OHE}}+\sigma_{\textrm{OHE}}^{\textrm{int}}$) in the presence of dilute random impurities for selected scalar potential strengths. Inset: Extrinsic OHC as a function  of $u_z/u_0$ for $\varepsilon=1.2 \Delta$ (note that $|\sigma_{\textrm{OHE}}^{\textrm{int}}|$ is shown as a solid line for comparison). Parameters: $v=10^{6}\,\mathrm{m/s}$, $\Delta=0.5\,\mathrm{eV}$, $n=10^{12}\,\mathrm{cm^{-2}}$, and $u_{0}=1\,\mathrm{eV}\,\mathrm{nm}^{2}$ (inset).}\label{fig:02}
		\par\end{centering}
\end{figure}

The first term of Eq. (\ref{eq:OHE_scalar}) is inversely proportional to the potential strength and  has odd parity, i.e. $\sigma_{\textrm{OHE}}^{\textrm{s.c.}}(\varepsilon)=-\sigma_{\textrm{OHE}}^{\textrm{s.c.}}(-\varepsilon)$, unlike $\sigma_{\textrm{OHE}}^{\textrm{int}}$ which is an even function of $\varepsilon$ \cite{Bhowal2021}. This symmetry is perturbatively broken due to disorder, and, for intermediate scattering strengths, an important next-order contribution [the second term in Eq.(\ref{eq:OHE_scalar})] kicks in. This effect can be seen in Fig. \ref{fig:02}, where the total OHC, $\sigma_{\textrm{OHE}}^{\text{tot}}=\sigma_{\textrm{OHE}}^{\text{dis.}}+\sigma_{\textrm{OHE}}^{\textrm{int.}}$, is plotted against the Fermi energy for two choices of $u_0$ and a fixed $n$. We note that  $\sigma_{\textrm{OHE}}^{\text{dis.}}$ matches well the analytical approximation of Eq. (\ref{eq:OHE_scalar}) in this low-$u_0$ regime. The extrinsic OHC is seen to depend strongly on $u_0$ and in the metallic regime it can \textit{easily exceed} $\sigma_{\textrm{OHE}}^{\textrm{int.}}$, especially for low $n$. To understand this behavior, one needs to pin down the exact underlying mechanism of the extrinsic response. The sensitivity of $\sigma_{\textrm{OHE}}^{\text{dis.}}$ to the impurity potential strength suggests that an orbital version of the familiar (spin-dependent) skew scattering mechanism is at play. To confirm this, we also perform a standard ladder resummation. The rationale is that, by construction, the ladder approximation excludes semiclassical skew-scattering diagrams (most notably the $Y$ diagram \cite{Milletari2016}). We find $\sigma_{\textrm{OHE}}^{\textrm{ladder}}=0$ (to the leading order in $n$), which confirms our hypothesis. Due to the semiclassical nature of the skew scattering mechanism, one has $\sigma_{\textrm{OHE}}\sim n^{-1}$  akin to the familiar Drude conductivity ($\sigma_{xx}$). However, the transport times governing each response function are drastically different. To leading order in $u_0$, one has $\sigma_{\textrm{OHE}}^{\text{s.c.}}(\varepsilon)\propto(n\varepsilon^{2}u_{0})^{-1}$, while $\sigma_{xx}(\varepsilon)\propto\varepsilon\tau_{\parallel}$ with $\tau_{\parallel}\propto(n\varepsilon u_{0}^{2})^{-1}$ (here, the high Fermi energy limit $\varepsilon \gg \Delta$ was taken for simplicity). This shows that, in analogy to the extrinsic SHE, the  \textit{orbital-Hall response is governed by its own transverse transport time}, namely, $\sigma^{\text{s.c.}}_{\textrm{OHE}}\propto v k_F \tau_{\perp}$, with $k_F=\sqrt{f_1(\varepsilon,\Delta)}$ [the  parametric dependencies of $\tau_{\perp}$ can be read off from Eq. (\ref{eq:OHE_scalar})]. Due to $\mathcal T$ symmetry, the skewness of impurity cross sections features a \textit{valley-orbit locking effect} akin to the intrinsic OHE mechanism \cite{Bhowal2021}; see Fig. \ref{fig:02} (top panel).  

The characteristic behavior of the OHC reflects the structure of the disorder-renormalized vertex  $\tilde{J}_{x}$. We find $\tilde{J}_{x}=a\,J_{x}+b\,J_{y}$, with $a,b$ some $\mathcal O (n^0)$ constants \cite{Supplementary_material}. The $\mathcal O (n^0)$ $J_y$ term (absent in the ladder approximation)  shows that, through skew scattering, disorder acts as a robust source of transverse OAM flow (note that $\mathcal{J}_{y}^{z} \propto J_y$ in our model). Moreover, the nonperturbative dependence of $\tilde{J}_{x}$ on the scattering potentials $u_0$ and $u_z$ is accessible via our technique \cite{Supplementary_material}. It is instructive to compare our findings to Ref. \cite{liu_dominance_2023}, where a white-noise (WN) model of scalar disorder was employed. There, the leading term of the small-$n$ expansion reads $\sigma^{\text{WN}}_{\textrm{OHE}}\propto (n)^{0}$, suggesting that a noncrossing calculation is insufficient (see Refs. \cite{Ado_2015,Milletari2016} for a discussion of the breakdown of perturbative analysis in 2D Dirac models with WN and similar zero-spatial-average disorder landscapes). The WN statistics yield an extrinsic OHC \textit{independent} on the disorder details, as well as an unphysical $C_0$-type discontinuity as the Fermi energy approaches the band edge \cite{liu_dominance_2023}. In contrast, in our model (of random short-range impurities), the extrinsic OHC shows regular behavior across the band edge and the semiclassical skew scattering mechanism is operative ($\sigma^{\text{s.c.}}_{\textrm{OHE}}\propto n^{-1}$), leading to a  physically sound $\sigma_{\textrm{OHE}}^{\text{dis.}}$ that is sensitive to $u_0$ and $n$ as expected in a realistic disordered material.

\begin{figure}
	\begin{centering}
        \includegraphics[scale=0.45]{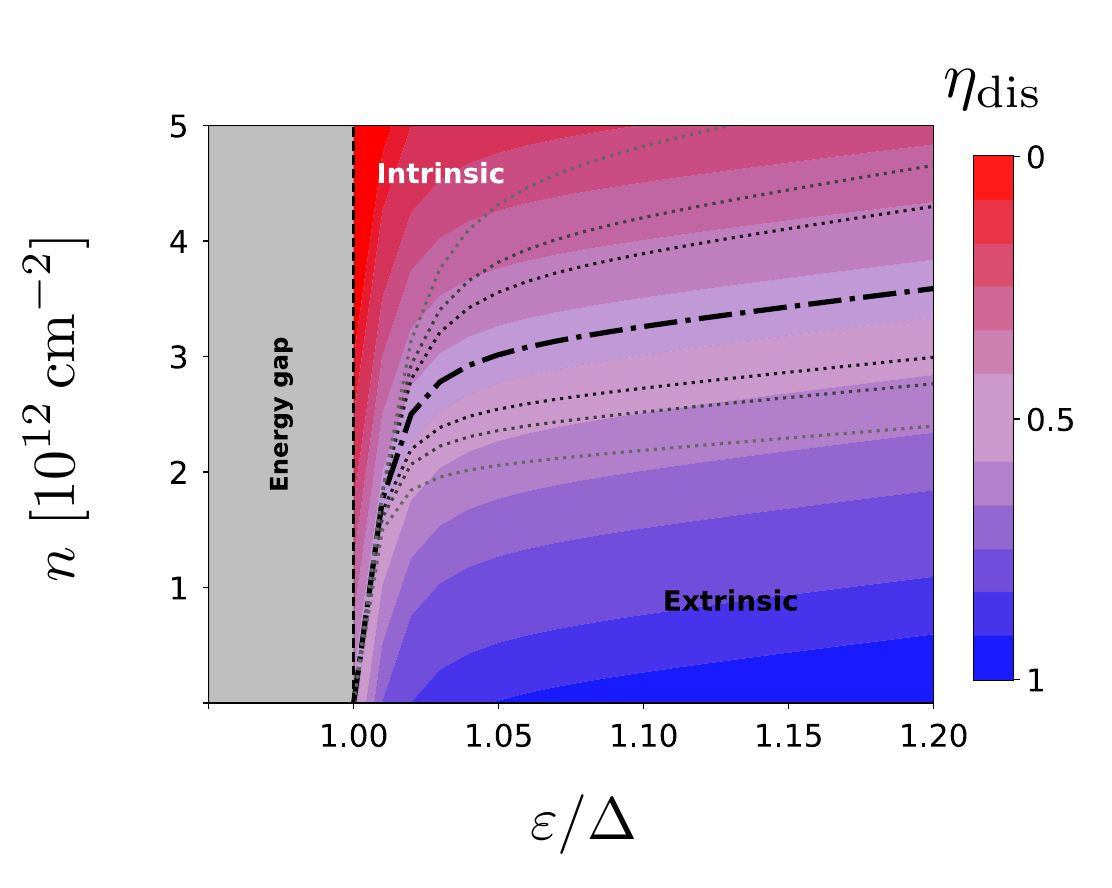}\caption{Crossover between the intrinsic and extrinsic regimes of the OHE in a system with scalar disorder. The dash-dotted curve corresponds to the critical line  $\eta_{\text{dis}}=1/2$ for impurities with $u_{0}\equiv u_{0}^*=100\,\mathrm{eV nm}^{2}$. The dotted lines show how the  boundary changes upon tuning $u_0$. These transition from the left to the right of the $u_0^*$ curve with $u_0$ increasing as $u_{0}/u_{0}^*=0.67,\,0.77,\,0.83,\,1,\,1.2,\,1.3,\,1.5$. The color map represents the relative extrinsic contribution strength $\eta_{\textrm{dis}} \in [0:1]$. Inside the energy gap, the OAM transport is governed by the orbital Berry curvature. Other parameters as in Fig. \ref{fig:02}.}
        \label{fig:03}
		\par\end{centering}
\end{figure}

It is natural to ask whether the intrinsic mechanism can prevail over orbital skew scattering in the regime of diffusive charge transport. To answer this question, we performed a detailed study in the parameter space spanned by $\varepsilon$ and the orbital mass of the gapped 2D Dirac model. Note that the low native defect concentrations \cite{Joucken_21} of graphene implies that, in this system (where a sizable $\Delta$ can be induced via strain fields \cite{Pablo_14,Jung_15}), skew scattering is expected to dominate the OAM transport. For this reason, we shall focus on the case of transition metal dichalcogenide  (TMD) monolayers. Here, the area density of point defects can reach a few $10^{13}$\,cm$^{-2}$ \cite{Qiu_13,Hong_15}, taking the system closer to the actual ``dirty limit''.  To establish a physical picture, we map out the relative contribution of the skew scattering-driven OHE, $\eta_{\textrm{dis}}\equiv\bar{\sigma}_{\textrm{OHE}}^{\text{dis}}/(\bar{\sigma}_{\textrm{OHE}}^{\text{dis}}+\bar{\sigma}_{\textrm{OHE}}^{\textrm{int}})$, where the bar denotes the absolute value and $\sigma_{\textrm{OHE}}$  is computed numerically to access the full nonperturbative regime. For very strong scalar potentials (characteristic of vacancy defects \cite{Ferreira2011}), we find that the OHE is essentially intrinsic provided $n \gtrsim 5 \times 10^{12}$ cm$^{-2}$. Note that TMD devices operating in the metallic regime (with $\varepsilon \approx  1.1\, \Delta$ \cite{Radisavljevic_13,Zhizhao_14}) have been demonstrated, so that pure intrinsic orbital transport (that is, $\eta_{\text{dis}}\ll 1$) may be within reach; see Fig. \ref{fig:03}. We checked that the side-jump contribution to the OHC is typically too low to overcome intrinsic orbital transport, especially for strong scattering potentials \cite{Supplementary_material}. Moreover, the extrinsic-to-intrinsic crossover is nonuniversal and smooth.  Similarly to charge transport  \footnote{Significant weak localization effects occur when the mean free path approaches the Fermi wavelength, $l = v \tau_0 \lesssim \lambda_F$. Due to the Dirac nature of the spectrum, this regime is restricted to a narrow energy window for systems with strong impurities unless $n$ is high (e.g. $\Delta < \epsilon \lesssim 1.1 \Delta$ for $u_0=u_0^*$ and $n \gtrsim 5\times 10^{12}$ cm$^{-2}$).}, the nature of OAM transport is sensitive to the carrier density and the details of the disorder model. As one moves away from the band edge into the metallic phase, the radius of the Fermi surface increases, which favors skew scattering processes. To better see this, in Fig. \ref{fig:03} we also show $n^*=n^*(\varepsilon,u_0)$, defined as the critical impurity density at which $\eta_{\text{dis}}=1/2$ and thus intrinsic and extrinsic mechanisms contribute equally. Below the critical $n$, the extrinsic OHC dominates. The family of dotted lines (which track the evolution of $n^*$ with $u_0$) disclose a more prevailing extrinsic behavior of the OAM transport for weaker scattering potentials, as well as for higher carrier densities. In contrast, intrinsic orbital transport is seen to dominate close to the band edge and deep inside the dirty limit. Due to the prevalence of atomic defects in TMDs \cite{Hong_15} with typical large $|u_0|$, these findings indicate that the two distinct OAM transport regimes should be accessible using a back-gate voltage. While this study is focused on zero-temperature properties, the general qualitative picture of the OHE remains the same at finite temperature, except for specific  regions of the parameter space where electron-phonon scattering plays a significant role (see Ref. \cite{Supplementary_material} for additional details).

The problem of extrinsic OAM transport becomes even more intriguing when considering non-scalar disorder. Scattering potentials endowed with a nontrivial internal structure in the unit cell are ubiquitous but have so far not been investigated in the context of the OHE. For case (ii), in the limit of a pure staggered potential (see \cite{Supplementary_material} for more general expressions), we find  
\begin{equation}
	\sigma_{\textrm{OHE}}^{\textrm{s.c.(z)}}=-\frac{2 e m_{e}v^{2}}{\pi n}\theta(|\varepsilon|-|\Delta|)\frac{\Delta\,}{u_{z}|\varepsilon|}\frac{f_{1}(\varepsilon,\Delta)}{f_{2}(\varepsilon,\Delta)}+\mathcal{O}(u_z^0).
 \label{eq:OHC_staggered}
\end{equation} 
For such impurities, the extrinsic OHC has instead even parity with respect to $\varepsilon$, which demonstrates that the \textit{spatial symmetries} of local scattering potentials can have a large impact on the OHE. Note that the scaling with $\varepsilon$ and $\Delta$ is also modified with respect to the case of scalar impurities [c.f. Eq. (\ref{eq:OHE_scalar})]. This is interesting, but challenging to probe experimentally as \textit{pure} staggered $\delta$-type potentials are not easily accessible \cite{Basko_08,Bena_08,Pachoud_14}. To see how the skewness of orbital scattering processes may vary between different types of common impurities, we investigate the OHC dependence on the ratio $u_z/u_0$. A representative study for $\varepsilon=1.2 \Delta$ is shown in the inset of Fig. \ref{fig:02}. Upon increasing the staggered component of the potential, we see a quick reduction of the OHC from about $\approx 145\, (e/2\pi)$ at $u_z=0$ ($C_{6v}$-symmetric impurities) to zero for $u_z=u_0$. The latter is the special case of short-ranged potentials localized on a single sublattice ($C_{3v}$-symmetric impurities), for which orbital skew scattering is conspicuously inactive.  For $u_z>u_0$, the OHC changes sign and develops a nonmonotonic behavior, highlighting a subtle competition between orbital skew scattering processes of distinct origin. This confirms the intuition developed through Eqs. (\ref{eq:OHE_scalar})-(\ref{eq:OHC_staggered}), that is, the structure of impurity potentials is key to developing a quantitative and qualitative description of extrinsic OAM transport. 

\textit{Conclusion}.---Taken together, our findings reveal a rich, hitherto unreported, interplay of OAM transport mechanisms that reflect the band structure and disorder landscape. In particular, we uncover an orbital analog of the familiar skew scattering mechanism, which is sensitive to the symmetry and strength of local impurity potentials. Despite our focus on 2D honeycomb layered systems with broken inversion symmetry, many of the OHE features we described here hold quite generally. Most importantly, the new orbital skew scattering mechanism is expected to be universal to orbital-active 2D electronic systems, both in the presence and absence of $\mathcal T$ symmetry. In particular, it will play a key role in centrosymmetric materials with hidden out-of-plane orbital textures, as shown in the Supplementary Material for the specific case of $D_{6h}$-invariant graphene with only intrinsic-type spin-orbit coupling \cite{Supplementary_material}. The prevalent manifestation of orbital skew scattering  is a crucial result and one of the main consequences of the microscopic OHE framework developed in this work.

\begin{acknowledgments}
T.G.R. acknowledges funding from  the Portuguese Foundation for Science and Technology (FCT)  through \\ Grant Nos. 2022.07471.CEECIND/CP1718/CT0001\\ (DOI identifier: 10.54499/2022.07471.CEECIND/\\CP1718/
CT0001)  and 2023.11755.PEX (DOI identifier: 10.54499/2023.11755.PEX) and  Strategic Funding UIDB/04650/2020. A.F. acknowledges partial support from the International Spintronics Network (Grant No. EP/V007211/1) funded by UK's Engineering and Physical Sciences Research Council. A. F. further thanks Henrique P. V. C. Jorge for a close reading of the manuscript and for useful comments on the accuracy of the analytical approximations for the orbital Hall conductivity.
\end{acknowledgments}

\begin{widetext}

%-------------------------
%% Supplemental Material    
%-------------------------

%\onecolumngrid

\section*{Supplementary Material}

%%%%%%%%%%%%%%%%%%%%%%%%%%%%%%%%%%%%%%%%%%%%%%%%%%%
%%%%%%%%%%%%%%%%%%%%%%%%%%%%%%%%%%%%%%%%%%%%%%%%%%%
\section*{Preliminaries}
%%%%%%%%%%%%%%%%%%%%%%%%%%%%%%%%%%%%%%%%%%%%%%%%%%%
%%%%%%%%%%%%%%%%%%%%%%%%%%%%%%%%%%%%%%%%%%%%%%%%%%%
Natural units with $\hbar,e \equiv 1$ are used in this supplementary material. Note that the basic definitions regarding the 2D Dirac Hamiltonian and the disorder potential created by $\delta$-type impurities can be found in the main text. We remind the reader that the Hamiltonian in the valley-isotropic basis $(AK,BK,-BK^\prime,A)$ is

\begin{equation}
H_\tau=H_{0,\tau}+V_{\text{dis},\tau}=v\,\boldsymbol{\sigma}\cdot\mathbf{p}+\tau \Delta\,\sigma_{z}+V_{\text{dis},\tau}\,
\label{eq:ham},
\end{equation}
where $V_{\text{dis},\tau}$ is the disorder term. Barring spin-orbit effects that are beyond the scope of the current work (required, for example, to describe valence-band splittings in group-VI transition metal dichalcogenides \cite{Kormanyos_2015}), the clean Hamiltonian $H_{0,\tau}$ approximates well the low energy properties of a range of honeycomb monolayers and graphene heterostructures, such as graphene-on-hBN \cite{Zollner_19}. Eq. (\ref{eq:ham}) is a well established model for studies of impurity-limited transport \cite{RevModPhys.82.2673,RevModPhys.83.407} that has enabled tractable microscopic theories with predictive power in various contexts including spintronics \cite{Perkins2024_review}. To simplify the notation, the metallic regime ($|\varepsilon| > |\Delta|$) is assumed in all expressions below.

%%%%%%%%%%%%%%%%%%%%%%%%%%%%%%%%%%%%%%%%%%%%%%%%%%%
%%%%%%%%%%%%%%%%%%%%%%%%%%%%%%%%%%%%%%%%%%%%%%%%%%%
\section*{Section I: T-matrix and disorder-averaged Green's functions}
%%%%%%%%%%%%%%%%%%%%%%%%%%%%%%%%%%%%%%%%%%%%%%%%%%%
%%%%%%%%%%%%%%%%%%%%%%%%%%%%%%%%%%%%%%%%%%%%%%%%%%%
The retarded($+$)/advanced($-$) disorder-averaged Green's function (GF) in $\textbf{k}$-space is
%%%%%%%%%%%%%%%%%%%%%%%%%%%%%%%%%%%%%%%%%%%%%%%%%%%
\begin{equation}
\mathcal{G}_{\varepsilon,\tau}^{\pm}(\mathbf{k})=(\varepsilon-H_{0,\tau}(\mathbf{k})-\Sigma^{\pm}_{\varepsilon,\tau})^{-1},  
\label{eq:GF_def}
\end{equation}
%%%%%%%%%%%%%%%%%%%%%%%%%%%%%%%%%%%%%%%%%%%%%%%%%%%
where $H_{0,\tau}(\textbf{k})$ is the bare Hamiltonian and $\Sigma^{\pm}_{\varepsilon,\tau}$ is the disorder self-energy. Because intervalley scattering is neglected, all operators are diagonal in the valley space. For ease of notation, we define $\Delta_\tau = \tau \Delta $.  Within the $T$-matrix approach, the full Born series of single-impurity scattering events is summed exactly, yielding $\Sigma_{\varepsilon,\tau}^{a}=nT^{a}_{\tau}(\varepsilon)$, where $T^{a}_{\tau}$ is the single-impurity $T$-matrix, $a=\pm1$ and $n$ is the impurity density. $T^{a}_{\tau}$ admits a closed form expression given by 
%%%%%%%%%%%%%%%%%%%%%%%%%%%%%%%%%%%%%%%%%%%%%%%%%%%
\begin{equation}
T^{a}_{\tau}(\varepsilon) = T_{0,\tau}^{a}(\varepsilon)\,\sigma_{0}+T_{z,\tau}^{a}(\varepsilon)\,\sigma_{z},
\label{eq:Tmatrix}
\end{equation}
where  
\begin{equation}
T_{0,\tau}^{a}(\varepsilon)=\sum_{p=\pm1}\frac{u_{0}+p \tau u_{z}}{2}\frac{1}{1-(pu_{0}+\tau u_{z})(\Delta_\tau+p\varepsilon)\Theta^{a}(\varepsilon)},
\label{eq:T0}
\end{equation}
\begin{equation}
T_{z,\tau}^{a}(\varepsilon)=\sum_{p=\pm1}\frac{pu_{0}+\tau  u_{z}}{2}\frac{1}{1-(pu_{0}+\tau u_{z})(\Delta_\tau+p\varepsilon)\Theta^{a}(\varepsilon)}\,,
\label{eq:T3}
 \end{equation}
and 
\begin{equation}
\Theta^{a}(\varepsilon)=-\frac{1}{4\pi v^2}\,\left(i\pi a\text{\ensuremath{\frac{\varepsilon}{|\varepsilon|}}+}\ln\frac{\Lambda^{2}-\varepsilon^{2}+\Delta^{2}}{\varepsilon^{2}-\Delta^{2}}\right).   
\end{equation}
%%%%%%%%%%%%%%%%%%%%%%%%%%%%%%%%%%%%%%%%%%%%%%%%%%%
We recall that $\Lambda \gg |\varepsilon|,|\Delta|$ is the ultraviolet cutoff used to regulate divergent $k$ integrals in the low-energy approach. $\Lambda\approx 10$ eV is a common choice in the literature, but the numerical value of $\Lambda$ has little impact on the main physical quantities of interest. For example, the orbital transport coefficients are independent of $\Lambda$ (see below). Some quantities can show a weak logarithmic dependence on $\Lambda$, like the charge conductivity in the strong scattering regime.  Moreover, owing to the 2D Dirac nature of the spectrum, scattering resonances occur near the Fermi level for sufficiently large $|u_{0,z}|$, leading to conspicuous energy dependencies in physical quantities. The resonant energies $\varepsilon_p$ are determined by the poles of the matrix $T$, that is, by solving $\text{Re\,}\{(p u_{0}+u_{z})(\Delta_{\tau}+p\varepsilon_p)\Theta^{a}(\varepsilon_p)\}=1$.

\vspace{2mm}

The disorder-averaged GF is obtained by plugging the $T$-matrix [Eq. (\ref{eq:Tmatrix})] into Eq. (\ref{eq:GF_def}), yielding:
%%%%%%%%%%%%%%%%%%%%%%%%%%%%%%%%%%%%%%%%%%%%%%%%%%%
\begin{equation}
\mathcal{G}_{\varepsilon,\tau}^{a}(\mathbf{k})=-\frac{vk_{x}\sigma_{x}+vk_{y}\sigma_{y}+\left[\varepsilon-nT_{0,\tau}^{a}(\varepsilon)\right]\sigma_{0}+\left[\Delta_\tau+nT_{z,\tau}^{a}(\varepsilon)\right]\sigma_{z}}{v^2 k^{2}+\left[\Delta_\tau+nT_{z,\tau}^{a}(\varepsilon)\right]^{2}-\left[\varepsilon-nT_{0,\tau}^{a}(\varepsilon)\right]^{2}}\,,
\label{eq:GF_expression}
\end{equation}
with $k=|\textbf{k}|$. We see that impurity potentials with a non-zero staggered component ($u_z \neq 0$) generate a $\sigma_z$ contribution to the self energy that  renormalizes the orbital mass term according to $\Delta \rightarrow \Delta + n\, \text{Re\,} T_z^a(\varepsilon)$.

\vspace{2mm}

The quasiparticle self-energy encodes the typical scattering times of the model according to:

\begin{equation}
-\textrm{Im}\,\Sigma_{\varepsilon}^{+}\equiv\frac{1}{2\tau_{0}(\varepsilon)}\,\sigma_{0}+\frac{1}{2\tau_{z}(\varepsilon)}\,\sigma_{z},\,\,\,\,\,\,\,\,\,\,\,\,\,\hfill[2\tau_{\alpha}(\varepsilon)]^{-1}=-n\,\textrm{Im}\,T_{\alpha}^{+}(\varepsilon)\,,\quad(\alpha=0,z),
\end{equation}
where $\tau_{0(z)}$ is the  momentum scattering time associated to the scalar (staggered) potential and  we omitted the valley index for brevity. Hereafter, operators are taken at the $K$ valley and a valley degeneracy factor ($g_v=2$) is introduced when necessary.

%%%%%%%%%%%%%%%%%%%%%%%%%%%%%%%%%%%%%%%%%%%%%%%%%%%
%%%%%%%%%%%%%%%%%%%%%%%%%%%%%%%%%%%%%%%%%%%%%%%%%%%

\section*{Section II: Disorder-renormalized vertex and orbital Hall conductivity}

\begin{figure}
	\begin{centering}
		\includegraphics[scale=0.67]{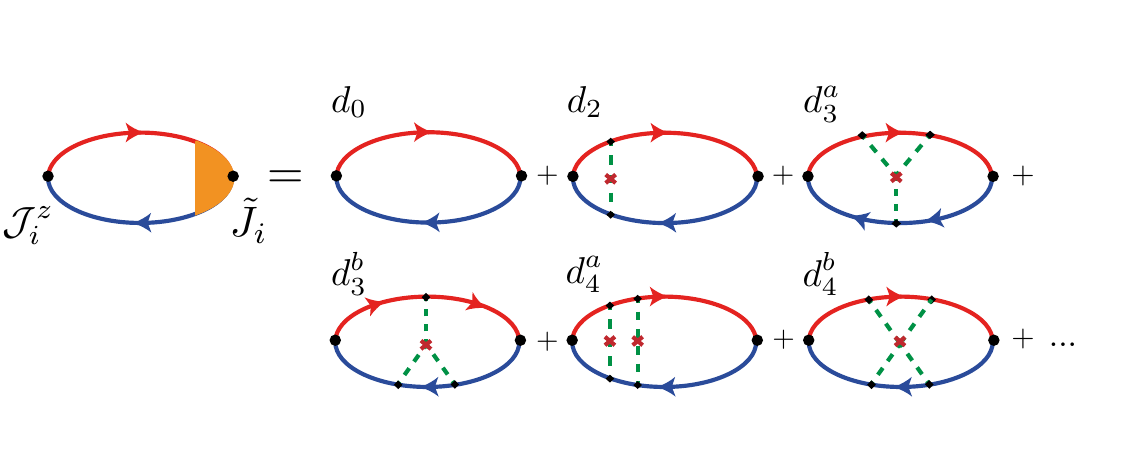}
    \caption{Infinite series of linear-response (2-particle GF) diagrams evaluated in this work using the $T$-matrix technique of Ref. \cite{Milletari2016} (only diagrams with up to 4  potential insertions are shown here). This is a nonperturbative approach at the single-impurity level that sums the complete series of noncrossing diagrams. In contrast, the popular ladder  technique  only captures the sub-class of diagrams with two-potential insertions (i.e., $d_2$ and $d_4^a$ in the figure), which is equivalent to a series of scattering events taken at the first-Born approximation level. The lowest-order skew scattering diagrams  are $d_3^a$ and $d_3^b$, dubbed the $Y$-diagrams, which are absent from the ladder series. The $T$-matrix resummation technique captures all these diagrams, and thus provide a virtually exact description of diffusive orbital transport that is valid for both weak and strong impurity scattering potentials insofar as quantum interference corrections (due to coherent scattering events involving 2 or more impurities \cite{Milletari2016})  are negligible.}
     \label{fig:supp_01}
		\par\end{centering}
\end{figure}

Our microscopic theory applies to diffusive systems with $\varepsilon \, \tau_0 \gg 1$, where $\tau_0 \propto 1/n$ (this is the typical scenario for high-quality 2D crystals). Impurity scattering events lead to the disorder self-energy (entering the single-particle GF) discussed previously, but are also responsible for the renormalization of the vertex function entering the linear-response (two-particle) GFs. Both effects are crucial for a correct description of the OHE.

We recall that the renormalized charge current vertex  is obtained by solving the $T$-matrix Bethe-Salpeter equation \cite{comment} (see  main text), formally given by
\begin{equation}
\tilde{J}_{x}=J_{x}+n\sum_{\mathbf{k}}\,T^{+}(\varepsilon)\mathcal{G}_{\varepsilon}^{+}(\mathbf{k})\tilde{J}_{x}\mathcal{G}_{\varepsilon}^{-}(\mathbf{k})T^{-}(\varepsilon),
\label{eq:BS_equation}
\end{equation}
where $J_x = -e v \sigma_x$ is the bare charge current vertex of the model. The insertion of $\tilde{J}_{x}$ on one side of the linear response bubble (the $\tilde J_x - \mathcal{J}_y^z$ bubble) yields the full series of noncrossing diagrams generated by an arbitrary number of single-impurity scattering events (see Fig. \ref{fig:supp_01}).

The exact solution of Eq. (\ref{eq:BS_equation}) is cumbersome, resulting in an unwieldy expression for $\tilde J_x$. However, power expansions can be used to yield insightful analytical results. In the presence of either scalar or magnetic impurities, and within the weak scattering limit ($|u_{0,z}(\Delta \pm \varepsilon) \Theta^+ | \ll 1 $), we find

\begin{equation}
\tilde{J}_{x}[u_0,u_z=0]\simeq \left(2-\frac{4\Delta^{2}}{\varepsilon^{2}+3\Delta^{2}}\right)\,J_{x}+\frac{\Delta u_{0}}{v^2}\frac{\varepsilon^{4}-\Delta^{4}}{(\varepsilon^{2}+3\Delta^{2})^2}\,J_{y}.
\label{eq:renormalized_Jx1}
\end{equation}

\begin{equation}
\tilde{J}_{x}[u_0=0,u_z]\simeq \left(2-\frac{4\varepsilon^{2}}{3\varepsilon^{2}+\Delta^{2}}\right)\,J_{x}-\frac{\varepsilon u_{z}}{v^2}\frac{\varepsilon^{4}-\Delta^{4}}{(3\varepsilon^{2}+\Delta^{2})^2}\,J_{y}.
\label{eq:renormalized_Jx2}
\end{equation}

The second term in the RHS of Eqs. (\ref{eq:renormalized_Jx1})-(\ref{eq:renormalized_Jx2}) (\textit{absent in the standard ladder approximation}) results from skew scattering. This shows that the disorder potential fundamentally modifies the $J_x$ vertex, giving rise to an additional matrix structure, $J_y \propto \sigma_y$. This occurs primarily through the $Y$ diagrams ($d_3^a$ and $d_3^b$ in Fig. \ref{fig:supp_01}) and is responsible for the Drude-like  scaling of the orbital Hall conductivity (OHC), $\sigma_{\text{OHE}} \propto 1/n$, reported in the main text. Note that the $J_y$ coefficient is \textit{linear} in the scattering potentials. We also see that at the band edge ($|\varepsilon|=|\Delta|$), the renormalization is trivial ($\tilde J_x \propto J_x$) leading to a vanishing extrinsic OHC, as expected due to the vanishing Fermi surface. 

\vspace{2mm}

In the strong-scattering regime, i.e. $|u_{0,z}(\Delta \pm \varepsilon) \Theta^+ | \gg 1 $, we find  

\begin{equation}
\tilde{J}_{x}[u_0,u_z=0]\simeq 2 J_{x} -\frac{16 \pi ^2 \Delta v^2}{u_0 \left(\Delta^2-\varepsilon ^2\right) \left(\Xi_{\varepsilon}+\pi ^2\right)} J_{y}.
\label{eq:renormalized_Jx1_SSL}
\end{equation}
\begin{equation}
\tilde{J}_{x}[u_0=0,u_z]\simeq 2 J_{x} + \frac{16 \pi ^2 \varepsilon v^2}{u_z \left(\Delta^2-\varepsilon ^2\right) \left(\Xi_{\varepsilon}+\pi ^2\right)} J_{y}.
\label{eq:renormalized_Jx2_SSL}
\end{equation}
where $\Xi_{\varepsilon}=\log^{2}\left(\frac{\Lambda^{2}}{\varepsilon^{2}-\Delta^{2}}\right)$. Similar to the weak-scattering limit, there is a $J_y$ term due to skew scattering. This time though the expression is manifestly non-perturbative and thus is not associated to a particular diagram. In the strict resonant limit ($|u_0|\rightarrow \infty)$, the skew scattering term is vanishing and thus so too is the extrinsic OHC.

\vspace{2mm}

The above considerations show that the symmetry and energy dependency of $\tilde J_x$ are key in determining the correct physical behavior of the extrinsic orbital response.  From Eqs. (\ref{eq:renormalized_Jx1})-(\ref{eq:renormalized_Jx2}), the expressions of the OHC reported in the main text can be obtained from the $\tilde J_x - \mathcal{J}_y^z$--bubble  via a final momentum integration:

\begin{equation}
\mathcal{\sigma}_{\textrm{OHE}}(\varepsilon)=g_{v}g_{s}\int d\mathbf{k}\,\textrm{tr}\,\left[\mathcal{J}_{y}^{z}(\mathbf{k})\,\varrho(\varepsilon,\mathbf{k})\right],\qquad\varrho(\varepsilon,\mathbf{k})=\frac{1}{2\pi}\mathcal{G}_{\varepsilon}^{+}(\mathbf{k})\,\tilde{J}_{x}\,\mathcal{G}_{\varepsilon}^{-}(\mathbf{k})\,.
\label{eq:OHC_expression}
\end{equation}
where $g_s=2$ is the spin degeneracy factor. Note that the \textit{momentum-resolved near-equilibrium density matrix} $\varrho(\varepsilon,\textbf{k})$ can be readily obtained from Eq. (\ref{eq:GF_expression}) and the renormalized vertex. The $\textbf{k}$-integral is evaluated by first  decomposing the density matrix $\varrho(\varepsilon,\textbf{k})$ in Eq. (\ref{eq:OHC_expression})  into the  Clifford algebra spanned by the $\sigma_i$ matrices ($i=0,x,y,z$), 
\begin{equation}
\label{eq:Cdecomposition}
\varrho(\varepsilon,\textbf{k})=\sum_{i=0,x,y,z} \varrho_i(\varepsilon,\textbf{k}) \, \sigma_i.
\end{equation}
Clearly, the extrinsic OHE requires that the linear-response density matrix $\varrho(\varepsilon, \textbf{k})$ acquires a $\sigma_y$ component; without such a component the trace over pseudospin in Eq. (\ref{eq:OHC_expression}) vanishes exactly \cite{comment2}. Relatively compact expressions  can be obtained by writing $\tilde{J}_{x} = \alpha_x(\varepsilon) J_x + \alpha_y(\varepsilon) J_y$. We obtain

\begin{align}
\varrho_0(\varepsilon,\textbf{k}) &= k v \big\{\sin \theta_{\mathbf{k}} \big[ 2 \alpha_{y}(\varepsilon) \varepsilon -\alpha_{y}(\varepsilon) n (T_0^a+T_0^r)-i \alpha_{x}(\varepsilon) n
   (T_z^a-T_z^r)\big]- \nonumber\\
   &\quad \cos \theta_{\mathbf{k}} \big[-2 \alpha_{x}(\varepsilon) \varepsilon +\alpha_{x}(\varepsilon) n (T_0^a+T_0^r)-i \alpha_{y}(\varepsilon) n
   (T_z^a-T_z^r)\big]\big\}/P_1(\varepsilon)   \\[10pt]
   \label{eq:rho_0}
\varrho_x(\varepsilon,\textbf{k}) &=\big\{ k^2 v^2 (\alpha_{x}(\varepsilon) \cos 2 \theta_{\mathbf{k}}+\alpha_{y} \sin 2 \theta_{\mathbf{k}})+n \big[n (\alpha_{x}(\varepsilon) T_0^a T_0^r+i
   \alpha_{y}(\varepsilon) T_0^a T_z^r-\nonumber\\
   &\quad i \alpha_{y}(\varepsilon) T_0^r T_z^a-\alpha_{x}(\varepsilon) T_z^a T_z^r)-\alpha_{x}(\varepsilon) \varepsilon  (T_0^a+T_0^r+T_z^a+T_z^r)+i \alpha_{y}(\varepsilon) \varepsilon 
   (T_0^a-T_0^r+T_z^a-T_z^r)\big]\big\}/P_1(\varepsilon) \\[10pt]
   \label{eq:rho_x}
\varrho_y(\varepsilon,\textbf{k}) &=\big\{k^2 v^2 (\alpha_{x}(\varepsilon) \sin 2 \theta_{\mathbf{k}}-\alpha_{y}(\varepsilon) \cos 2 \theta_{\mathbf{k}})+n \big[ n (\alpha_{y}(\varepsilon) T_0^a T_0^r-i
   \alpha_{x}(\varepsilon) T_0^a T_z^r+\nonumber\\
   &\quad i \alpha_{x}(\varepsilon) T_0^r T_z^a-\alpha_{y}(\varepsilon) T_z^a T_z^r)+i
   \alpha_{x}(\varepsilon) \varepsilon  (-T_0^a+T_0^r-T_z^a+T_z^r)-\alpha_{y}(\varepsilon) \varepsilon 
(T_0^a+T_0^r+T_z^a+T_z^r) \big]\big\}/P_1(\varepsilon)  \\[10pt]
    %\label{eq:rho_y}
\varrho_z(\varepsilon,\textbf{k}) &=k v \big\{\sin \theta_{\mathbf{k}} \big[2 \alpha_{y}(\varepsilon) \varepsilon +i \alpha_{x}(\varepsilon) n (T_0^a-T_0^r)+\alpha_{y}(\varepsilon) n (T_z^a+T_z^r)\big]+\nonumber\\
&\quad \cos \theta_{\mathbf{k}} \big[ 2 \alpha_{x}(\varepsilon) \varepsilon -i \alpha_{y}(\varepsilon) n (T_0^a-T_0^r)+\alpha_{x}(\varepsilon) n
   (T_z^a+T_z^r)\big]  \big\}/P_1(\varepsilon)
    \label{eq:rho_z}
\end{align}
 where $\theta_{\textbf{k}}$ is the wavevector angle, $P_1 (\varepsilon) =\pi  D_A(\varepsilon) D_R(\varepsilon) $, and $D_{A(R)}(\varepsilon)$ is the denominator of the disorder-averaged advanced (retarded) GF in Eq. (\ref{eq:GF_expression}). We note that the above expressions for $\varrho_i(\varepsilon,\textbf{k})$ can be used to compute the orbital Hall response in any scattering regime, provided that the amplitudes of the renormalized vertex, $\alpha_{x,y}(\varepsilon)$, are known analytically (or determined numerically).

\vspace{2mm}

%%%%%%%%%%%%%%%%%%%%%%%%%%%%%%%%%%%%%%%%%%%%%%%%%%%
%%%%%%%%%%%%%%%%%%%%%%%%%%%%%%%%%%%%%%%%%%%%%%%%%%%

With the framework now in place, several useful analytical expressions can be obtained, including the semiclassical OHC reported in the main text. The usual small $n$-expansionsin the weak scattering limit (WSL) [c.f. Eq. (\ref{eq:renormalized_Jx1})-(\ref{eq:renormalized_Jx2})] for mixed impurities with $u_0\gg u_z$ (or $u_z \gg u_0$)  yield the following supplementary analytical results
\begin{equation}  
\left.\sigma_{\textrm{OHE},\text{WSL}}\right|_{u_{0}\gg u_{z}}\simeq\left.\sigma_{\textrm{OHE},\text{WSL}}\right|_{u_{z}=0}-\frac{2m_{e}v^{2}\Delta u_{z}\left(\varepsilon^{2}-\Delta^{2}\right)^{2}\left(13\Delta^{2}-\varepsilon^{2}\right)}{\pi nu_{0}^{2}|\varepsilon|\left(\varepsilon^{2}+3\Delta^{2}\right)^{3}}+\mathcal{O}(n^{0}),
\label{eq:OHC_1}
\end{equation}
\begin{equation}
\left.\sigma_{\textrm{OHE},\text{WSL}}\right|_{u_{z}\gg u_{0}}\simeq\left.\sigma_{\textrm{OHE},\text{WSL}}\right|_{u_{0}=0}-\chi_{\varepsilon}\frac{2m_{e}v^{2}u_{0}\Delta^{2}\left(\varepsilon^{2}-\Delta^{2}\right)^{2}\left(\Delta^{2}-13\varepsilon^{2}\right)}{\pi nu_{z}^{2}\varepsilon^{2}\left(3\varepsilon^{2}+\Delta^{2}\right)^{3}}+\mathcal{O}(n^{0}), 
\label{eq:OHC_2}
\end{equation}
with $\left.\sigma_{\textrm{OHE},\text{WSL}}\right|_{u_{z}=0}$ and $\left.\sigma_{\textrm{OHE},\text{WSL}}\right|_{u_{0}=0}$ given by Eqs. (\ref{eq:OHE_scalar})-(\ref{eq:OHC_staggered}) in the main text, respectively. Beyond the WSL, the expressions for the semiclassical contribution ($\mathcal{O}(n^{-1})$) for generalized impurity potentials with $u_0,u_z\neq0$ quickly become impractical, but numerical methods can be helpful, as shown in Sec. IV. 

\subsection*{Pure scalar and staggered impurities: orbital Hall angle in the weak and strong scattering limits}

Next, we compute the orbital Hall angle, defined as $\theta_{\text{oH}}=\sigma_{\text{OHE}}^{\text{s.c.}}/\sigma_0$, with $\sigma_0$ the charge conductivity and $\sigma_{\text{OHE}}^{\text{s.c.}}$ the semiclassical piece of $\sigma_{\text{OHE}}$. The charge conductivity, $\sigma_{0} = \sigma_{xx} = \sigma_{yy}$, can be written similarly to Eq. (\ref{eq:OHC_expression}), with the orbital current vertex replaced by $J_x$. In the vein of the main text, we will independently consider scalar ($p=0$) and ($p=z$) staggered impurities in the following. 

In the WSL, the Drude ($\mathcal{O}(n^{-1}$)) conductivity is

\begin{equation}
\sigma_{0,\text{WSL}}(\varepsilon)\simeq\frac{4v^{2}}{\pi nu_{p}^{2}}\frac{\varepsilon^{2}-\Delta^{2}}{g_{p}(\varepsilon)},
\label{weak_conductivity}
\end{equation}
where $g_0(\varepsilon)=\left(3 \Delta^2+\varepsilon ^2\right)$ and $g_z(\varepsilon)=\left( \Delta^2+3\varepsilon ^2\right)$.
In the strong scattering limit (SSL), on the other hand, we have
\begin{equation}
\sigma_{0,\text{SSL}}(\varepsilon)  \simeq \frac{(\varepsilon^{2}-\Delta^{2})\left(\Xi_{\varepsilon}+\pi^{2}\right)}{4\pi^{3}v^{2}n},
\end{equation}
for either type of impurity (scalar or staggered).
\vspace{2mm}
Combining Eq. (\ref{weak_conductivity}) with Eq. (3) in the main text, we can derive an analytic expression for the orbital Hall angle, which estimates the conversion efficiency from charge to orbital current. In the WSL, it takes the form
\begin{equation}
\theta_{\mathrm{oH},\text{WSL}}^{[0(z)]}(\varepsilon) \simeq \pm\frac{\Delta^{2}u_{0(z)}m_{e}}{2\varepsilon^{2}}\frac{\varepsilon^{2}-\Delta^{2}}{g_{0(z)}(\varepsilon)},
\end{equation}
and in the SSL, we find
\begin{equation}
\theta_{\mathrm{OH},\text{SSL}}^{[0(z)]}(\varepsilon)   \simeq  \pm  \frac{8m_{e}\pi^{2}}{u_{0(z)}\left(\varepsilon^{2}-\Delta^{2}\right)\left(\Xi_{\varepsilon}+\pi^{2}\right)}h_{0(z)}(\varepsilon) ,
\end{equation}
where $h_{0}(\varepsilon) = \Delta^2 / \varepsilon^2$, $h_{z}(\varepsilon)  = \Delta / \varepsilon$. Here, we used the following OHC expression in the SSL
\begin{equation}
   \sigma_{\mathrm{oH, \mathrm{SSL}}}^{ \text{s.c.};[0(z)]}(\varepsilon) \simeq \pm \frac{2 m_e \Delta^2 v^2}{\pi  n u_{0(z)} \varepsilon ^2}.
   \label{eq:OHC_SSL}
\end{equation}

An important remark on the significance of these results is in order. The OHC in our theory shows a clear dependence on $u_0$ (and $n$). However, this expected behavior of extrinsic transport is absent from the OHC computed for models of white-noise disorder in previous work \cite{liu_dominance_2023,Tang2024}, an artifact produced both by the assumption of white-noise statistics and by the use of perturbative disorder averaging techniques. In our realistic model of random impurities, $V_{\text{dis}}(\mathbf{x})=\sum_{i} M_{\textrm{dis}}(u_0,u_z)\,\delta(\mathbf{x}-\mathbf{x}_{i})$, such an artifact can be reproduced by carrying out non-self-consistent calculations at the first-Born approximation level (i.e. by keeping only ladder diagrams). Such an scheme of course not only misses skew-scattering physics that requires going beyond the first Born series [see, e.g. Eq. (\ref{eq:OHC_SSL}) computed non-perturbatively], but also misses the $u_0$ dependence of quantum (side jump) corrections as we will see next.

\section*{Section III: Orbital quantum-side jump}

\begin{figure}[h]  
\includegraphics[width=0.4\linewidth]{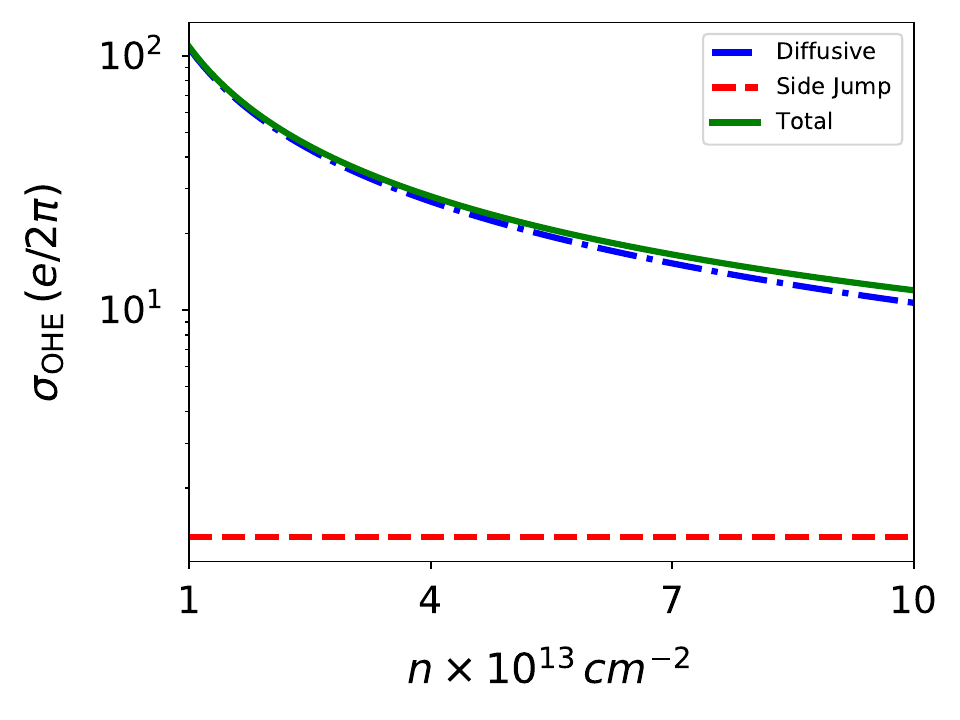}  \includegraphics[width=0.4\linewidth]{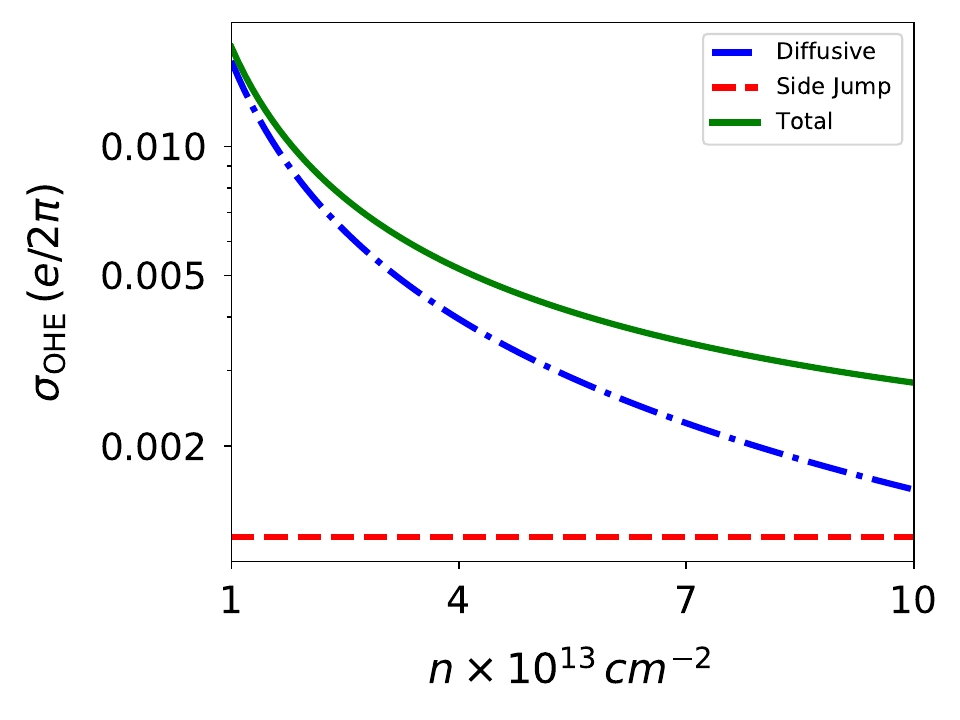}
\caption{Impurity density evolution of the OHC in the weak (left) and strong (right) scattering limits. The dot-dashed blue line  shows the  skew-scattering   OHC. The side-jump correction is shown in the dashed curves (red) and the total OHC in the solid (green) curves.  Parameters: $\varepsilon=1.2 \Delta$, $\Delta=0.5$ eV, $v=10^6$ m/s, $u_0=0.1$ eV nm$^2$ (left) and $u_0=100$ eV nm$^2$ (right).}
\label{fig:supp_fig_02}
\end{figure}

We now address the quantum side-jump correction to the OHC. Our technique allows to compute the side-jump correction to all orders in the scattering potential, unlike the standard wavepacket \cite{Sinitsyn_06} and quantum kinetic approaches \cite{liu_dominance_2023,Tang2024} which are limited to a description at the level of the first Born approximation. The  side jump effect is responsible for an important $\mathcal O(n^0)$ term in the small-$n$ expansion of the OHC:
\begin{equation}
\sigma_{\textrm{OHE}}(\varepsilon)=\underset{\sigma_{\textrm{OHE}}^{\text{s.c.}}}{\underbrace{S(\varepsilon)/n}}+Q(\varepsilon,n)\equiv\sigma_{\textrm{OHE}}^{\text{s.c.}}(\varepsilon)+\left[\sigma_{\textrm{OHE}}^{\text{s.j.}}(\varepsilon)+\sigma_{\textrm{OHE}}^{\text{coh}}(\varepsilon)\right]+\mathcal{O}(n)\,,  
\end{equation}
which, like  $\sigma_{\textrm{OHE}}^{\text{s.c.}}(\varepsilon)$, results exclusively from non-crossing diagrams. The additional $\mathcal O(n^0)$ contribution inside brackets, i.e. the term $\sigma_{\textrm{OHE}}^{\text{coh}}(\varepsilon)$, is due to crossing diagrams, which physically correspond to multiple coherent skew scattering events involving 2 or more impurities  \cite{Milletari2016}, and is not considered here.

\vspace{2mm}

The evaluation of the side-jump (SJ) contribution to the OHC requires a next-leading-order expansion of Eq. (\ref{eq:OHC_expression}) in impurity concentration $n$. This can be done with the help of Eqs. (\ref{eq:rho_0})--(\ref{eq:rho_z}) (formally exact), with $\alpha_{x,y}$--coefficients taken to   $\mathcal O (n)$. For illustration purposes, we consider only the case of scalar impurities. Evaluating Eq. (\ref{eq:OHC_expression}) to order $n^0$, we find
\begin{equation}
\sigma^{\mathrm{s.j.}}_{\text{OHE},\text{WSL}}  \simeq  \frac{4 m_e v^2 \Delta^2}{\pi |\varepsilon| (  3 \Delta^2  +  \varepsilon ^2)}
\end{equation}
in the  WSL and 
\begin{equation}
\sigma_{\text{OHE},\text{SSL}}^{\mathrm{s.j.}} \simeq - \frac{16m_{e}v^{4}\Delta^{2}\left[\varepsilon^{2}-\left(\varepsilon^{2}-\Delta^{2}\right)\log\left(\frac{\Lambda^{2}}{\Delta^{2}}\right)-\Delta^{2}\sqrt{\Xi_{\varepsilon}}\right]}{u_{0}\varepsilon^{4}\left(\varepsilon^{2}-\Delta^{2}\right)\left(\Xi_{\varepsilon}+\pi^{2}\right)}
\end{equation}
in the SSL, where $\Xi_{\varepsilon}=\log^{2}\left(\frac{\Lambda^{2}}{\varepsilon^{2}-\Delta^{2}}\right)$ as before. It is useful to represent these results graphically. In Fig. \ref{fig:supp_fig_02}, we show the $n$ dependence of $\sigma_{\textrm{OHE}}^{\text{s.j.}}(\varepsilon)$ for a Fermi energy of $\varepsilon = 1.2 \Delta$ in the two limiting regimes, the WSL and the SSL. We note that these results were carefully checked against numerics. In either scattering regime, the SJ contribution is relatively weak, except at very high impurity densities approaching $10^{14} $ cm$^{-2}$ (note that sufficiently strong scattering potentials are also required so that side jumps can compete against skew scattering). Such impurity densities exceed the typical density of native scatterers in 2D materials by up to 5 orders of magnitude \cite{Joucken_21,Hong_15}. In samples that are intentionally doped with ad-atoms, $n$ can be as high as $10^{12}$--$10^{13}$ cm$^{-2}$ \cite{RevModPhys.82.2673}. Hence, in adatom-engineered materials, the side jump contribution may compete with the intrinsic transport mechanism, and possibly become the dominant mechanism close to the band edge (very low Fermi energies), where skew scattering is suppressed. These supplementary results confirm that the hitherto-neglected \cite{liu_dominance_2023,Tang2024} orbital skew scattering processes dominate the extrinsic contribution to the OHC in most realistic scenarios (i.e. low $n$ and not too low $\varepsilon$). 

\section*{Section IV: On the accuracy of the analytical expressions in the WSL}

WSL series expansions are accurate up to intermediate scattering strengths. As such, our compact OHC expressions capture the semiclassical physics ($\varepsilon \tau_0 \gg 1$) over a significant range of parameters. To illustrate this, Fig. \ref{fig:supp_03} shows how the analytical OHC expression for scalar impurities [Eq. (3) in main text] fares against exact numerics. To this end, we evaluate the $\textbf{k}$-space integral in Eq. (\ref{eq:OHC_expression}) using the  full $\tilde J_x$ obtained via exact inversion of the $T$-matrix level Bethe-Salpeter equation [Eq. (\ref{eq:BS_equation})]. This required the development of a specialized numerical technique to be reported elsewhere. Figure \ref{fig:supp_03} shows that the next-leading-order correction [i.e. the $\mathcal{O}(u_{0}^0)$ term in the RHS of Eq. (3) in main text] is essential for an accurate description. As a check, we verified that our WSL expressions are valid for $|u_0|$  up to $0.2$ eV\,nm$^2$. For higher values of $|u_0|$, the quality of the $u_0$-expanded OHC expression gradually decreases, and a numerical evaluation becomes essential. Interestingly, in the unitary scattering limit ($|u_{0,z}|\rightarrow\infty$, relevant to point defects and resonant impurities \cite{RevModPhys.82.2673}) an analytical treatment is again feasible; see SSL expressions in Sec. II.

\vspace{3mm}

\begin{figure}[hbt!] 
	\begin{centering}
		\includegraphics[scale=0.5]{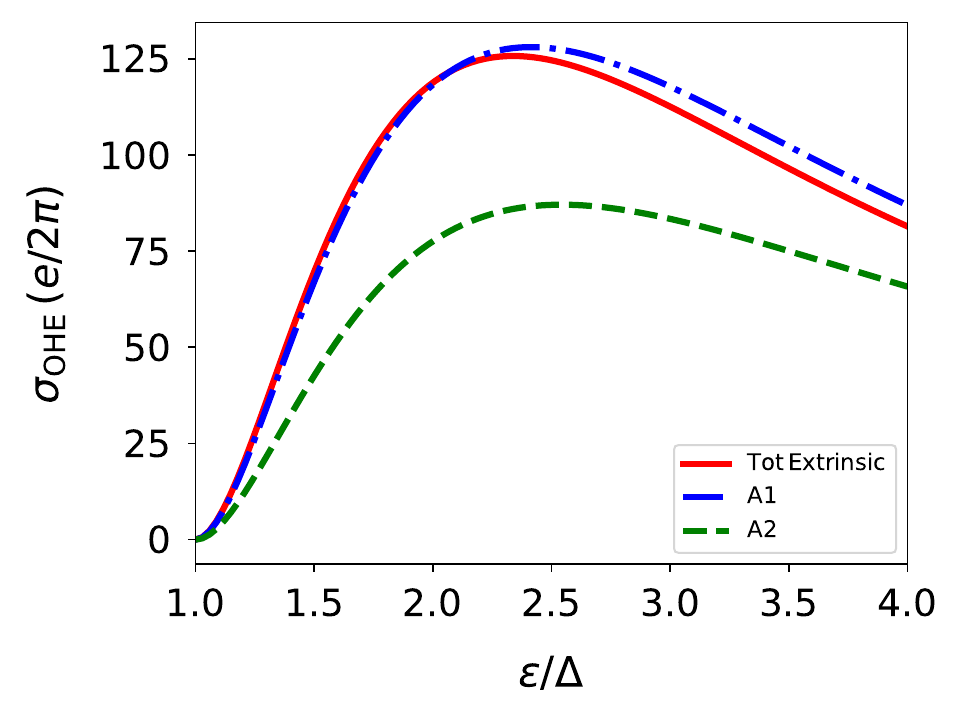}\caption{Fermi energy dependence of the OHC in the metallic regime obtained via exact numerical evaluation (red line) and analytical power series expansions in $u_0$ (dashed and dash-dotted lines). The lowest order term $\sim 1/u_{0}$ [see Eq. (3), main text] produces the dashed green line (labeled A2), while the inclusion of the next-order correction   returns the dash-dotted blue curve (labeled A1). Parameters: $v=10^{6}\,\mathrm{m/s}$, $\Delta=0.5\,\mathrm{eV}$, $n=10^{13}\,\mathrm{cm^{-2}}$, and $u_0 = 0.15\,\mathrm{eV}\mathrm{nm^2}.$ }
    \label{fig:supp_03}   
	\par\end{centering}
\end{figure}

\section*{Section V: Thermal excitations and electron-phonon scattering}

\begin{figure}[h]  
\includegraphics[width=0.4\linewidth]{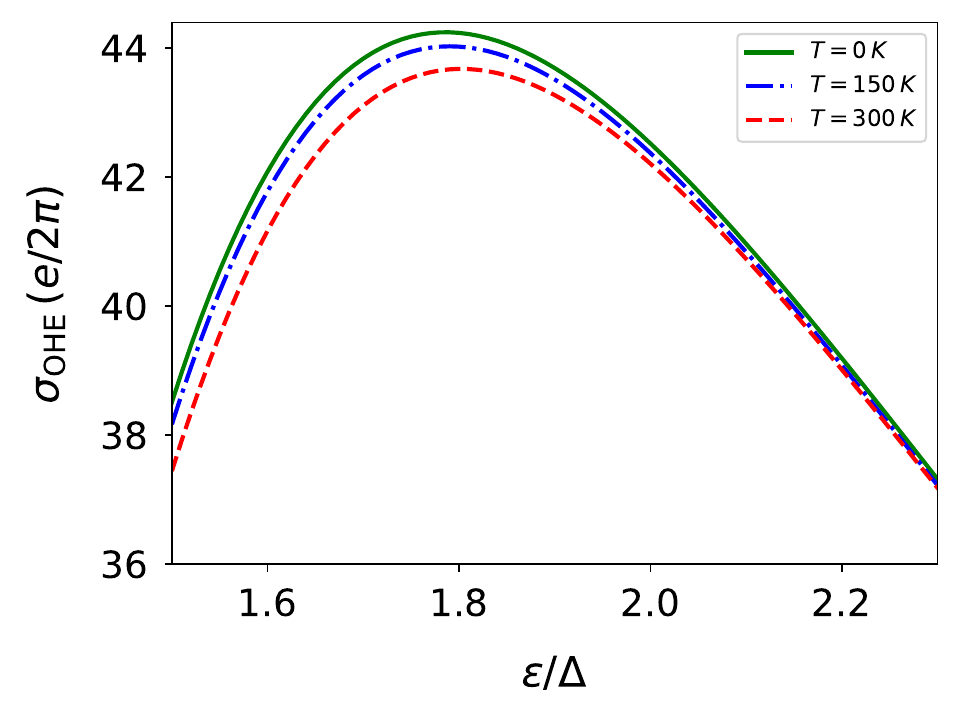}  \includegraphics[width=0.4\linewidth]{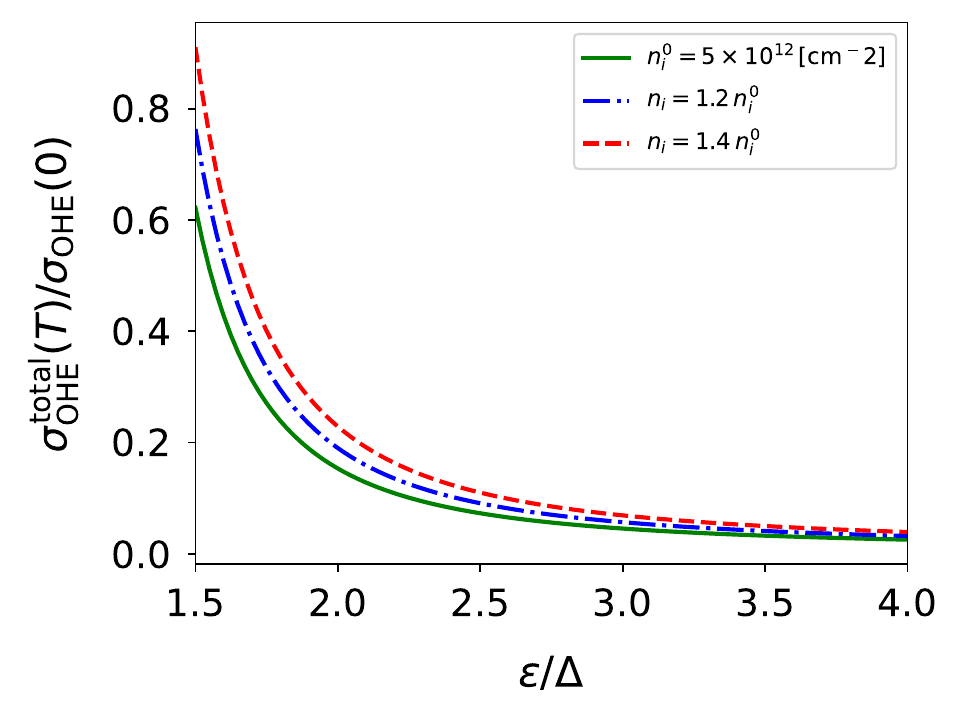}
\caption{\textbf{Left:} Extrinsic OHC near a local maximum at selected temperatures ($T=0, 150$ and $300$ K). Thermal fluctuations lead to a reduction of the OHC as the Fermi level approaches the band edge. Parameters as in Fig. 2 (main text). \textbf{Right:} Ratio of total OHC ($\sigma_{\text{OHE}}^{\text{total}}(T)$) to the purely extrinsic OHC ($\sigma_{\text{OHE}}(T)\approx \sigma_{\text{OHE}}(0)$) of gapped graphene at room temperature for selected impurity densities. Disorder parameters are chosen to tune the system into the phonon-dominated transport regime. Parameters: $v=10^{6}\,\mathrm{m/s}$, $\Delta=50$ meV, $u_0 = 0.1$ eV$\cdot$nm$^{2}$ and $T=300$ K (see text for phonon-related parameters).}
\label{fig:supp_fig_04}
\end{figure}

The diagrammatic formalism can be used to extend the theory beyond $T=0$ at various levels of approximation. The main finite-temperature effects expected to influence the OHE are: (i) thermal excitation of carriers at low electronic density and (ii) enhanced electron-phonon scattering processes. 

\vspace{2mm}
We first discuss the impact of thermal fluctuations. This mechanism is exponentially suppressed for $|\varepsilon| \gg k_B T$, but plays a role at low electronic densities. The left panel of Fig. \ref{fig:supp_fig_04} shows how the OHC reported in Fig. 2 of main text varies with $T$. We see that, as expected, lowering the Fermi level leads to a small reduction in the OHC. The weak $T$ dependence of the extrinsic OHC seen here can be traced back to the slow dependence of the transport coefficients with energy [see, e.g. Eq. (3) main text], which results in thermal averages very close to the $T=0$ values. Specifically, large energy fluctuations ($k_B T$) on the order of the orbital gap would be required to drastically reduce the OHC. 
\vspace{2mm}

Next, we address the role of electron-phonon scattering. We focus on the dilute impurity regime, where the dominant contribution to the $T=0$ OHC derives from Fermi-surface orbital skew scattering processes. We also specialize to gapped graphene systems, where orbital gaps $E_{\text{gap}}=2|\Delta|$ in the 10--50 meV range have been reported (see Ref. \cite{Jung_15}). To capture electron-phonon scattering effects, we add the the phonon self-energy $\Sigma_{\text{ph}}^a(\varepsilon,T)$ to the Green's functions using the non-crossing approximation, i.e.
\begin{equation}
\Sigma^a(\varepsilon,T)=\Sigma_{\text{ph}}^a(\varepsilon,T)+\Sigma_{\text{imp}}^a(\varepsilon)\,,
\label{eq:n.c.approximation_self_energy}
\end{equation}
where $\Sigma^a_{\text{imp}}(\varepsilon)\equiv \Sigma_{\varepsilon}^a$ is the   disorder self energy. Disorder averages are evaluated as done previously, whereas vertex corrections due to electron-phonon scattering are neglected. In the parameter region of interest, thermal fluctuations may be safely neglected. As such, we work with the $T=0$ formalism, while retaining the phonon effects via $\Sigma^a(\varepsilon,T)$. We focus on the interesting regime where the electron-phonon scattering rate exceeds the impurity contribution, giving way to strong $T$-dependencies in the OHC. Such a regime is expected to emerge in ultra-clean systems (i.e., very weak scattering potentials or very low $n$). We model the self-energy as $\Sigma_{\text{ph}}^a(\varepsilon,T) = - a i / ( 2\tau_\text{ph})$, where $\tau_\text{ph}$ is the scattering time due to longitudinal acoustic phonons in graphene. In the high-temperature limit, one has \cite{RevModPhys.83.407} 

\begin{equation}
\Sigma_{\text{ph}}^{\pm}(\varepsilon,T)\simeq\mp i\,\frac{\varepsilon}{2\hbar^{2}v^{2}}\frac{D^{2}}{\rho_{m}v_{\text{ph}}^{2}}\, k_{B}T,    
\label{eq:phonon_self_energy}
\end{equation}
where $D\approx 5$ eV is  deformation potential energy, $\rho_m\approx 7.6 \times 10^{-7}$ Kg$/$m$^{2}$ is the  mass density, and $v_{\text{ph}}\approx 2\times 10^4$ m$/$s is the sound velocity. The calculations carried out in this (phonon-limited) regime are summarized in the right panel of Fig. \ref{fig:supp_fig_04}. The main effect of electron-phonon scattering is a monotonic decrease in the total OHC with temperature \textit{and} Fermi level. Physically, this results from a substantial reduction of the orbital skew-scattering transport time ($\tau_\perp$) as the electronic density increases and electron-phonon processes become more  and more dominant; see Eq. (\ref{eq:phonon_self_energy}). 

\section*{Section VI: Universality of the orbital skew scattering mechanism: role of crystalline and time-reversal symmetries}

We now argue very generally that the basic requirement for orbital skew scattering (OSS) to manifest itself is an OMM-active band structure, that is, one in which a non-zero $L_z(\textbf{k})$ emerges for a fixed spin projection. This is true even in systems where the equilibrium OMM vanishes due to time-reversal symmetry, $\mathcal T$, but a hidden orbital texture is present (e.g. due to spin-orbit coupling). Remarkably, OSS is also operative (and thus will typically dominate the extrinsic OHE for $\varepsilon \tau_0 \gg 1$ or $n\ll n_e$, with $n_e$ the electronic density)  when $\hat H_0$ enjoys both $\mathcal T$ and inversion ($\mathcal I$) symmetries. To substantiate our claim, we now consider the extension of our microscopic theory to the paradigmatic Kane-Mele (intact $\mathcal T$ and $\mathcal I$) and Haldane (broken $\mathcal T$) models. These differ from the massive 2D Dirac model of Eq. (\ref{eq:ham}), which has intact $\mathcal T$ and broken $\mathcal I$, and will allow us to draw some general conclusions. 

\begin{figure}
	\begin{centering}
		\includegraphics[scale=0.75]{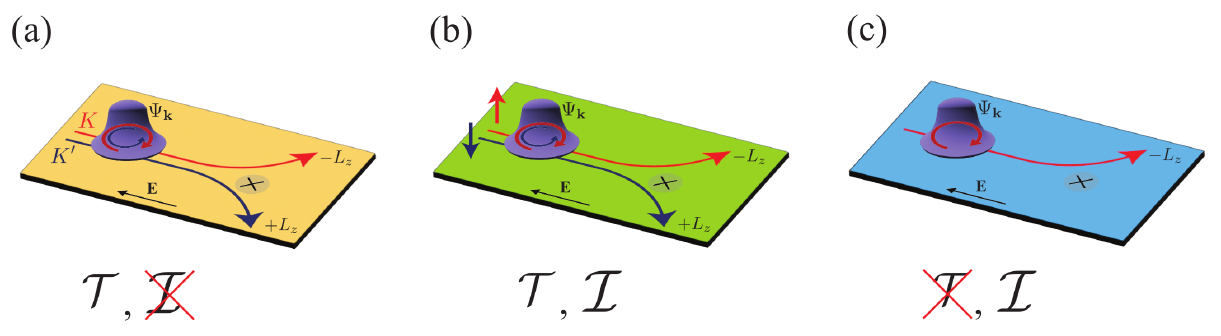}\caption{Depiction of diffusive OHEs governed by orbital skew scattering in 3 symmetry-distinct scenarios: \textbf{(a)} massive Dirac band model of graphene-on-hBN heterostructures and honeycomb monolayers with broken inversion symmetry; (\textbf{b}) Kane-Mele model of graphene with intrinsic spin-orbit coupling; and \textbf{(c)} 2D topogological (Chern) ferromagnet in the metallic phase.}
    \label{supp_fig_05}   
	\par\end{centering}
\end{figure}

\vspace{2mm}

We start with the Kane-Mele model. In the valley-isotropic basis, the Hamiltonian is

\begin{equation}
H_{\tau}^{(\textrm{KM})}=H_{0,\tau}^{(\textrm{KM})}+V^0_{\text{dis},\tau}=v\,\boldsymbol{\sigma}\cdot\mathbf{p}\,s_0+\lambda_{\text{KM}}\,\sigma_{z}s_{z}+\sum_{i}\,u_{0}\,\sigma_{0}s_{0}\,\delta(\mathbf{x}-\mathbf{x}_{i}),
\label{eq:ham_KM}
\end{equation}
where $\lambda_{\text{KM}}$ is the intrinsic-type spin-orbit coupling strength, $s_z$ is the diagonal Pauli matrix acting on the spin degree of freedom and the last term describes a conventional scalar disorder landscape. Note that $\hat H^{(\text{KM})}_{0}$ is invariant under $\mathcal{T}$  and $\mathcal{I}$. In fact, $\hat H^{(\text{KM})}_{0}$ is invariant under all operations of the $D_{6h}$ point-group describing flat graphene. Clearly, the Kane-Mele model corresponds to two copies of Eq. (\ref{eq:ham}) where the mass term sign depends upon the spin projection, i.e., we can map the two models by letting $\Delta \rightarrow \pm \tau \lambda_{\text{KM}}$ with the sign $\pm$ indicating the spin quantum number. From here, we immediately see that the OMM operator in the valley-isotropic basis is 

\begin{equation}
L_{z,\tau}^{\text{KM}}(\mathbf{k})=-\frac{\lambda_{\text{KM}}m_{e}v^{2}}{v^{2}k^{2}+\lambda_{\text{KM}}^{2}}\sigma_{0}s_{z}.
\label{eq:Lz_KM}    
\end{equation}

This shows that a hidden orbital texture emerges on each valley, $\mathcal L_{z,\tau,s}^{\text{KM}}(\textbf{k})=\langle \sigma s | L_{z,\tau}^{\text{KM}}(\mathbf{k}) | \sigma s\rangle$, with $s=\uparrow,\downarrow$ the spin quantum number. (Note that the OMM vanishes upon summing over the spin states because $\mathcal L_{z,\tau,s}^{\text{KM}}(\textbf{k})= -\mathcal L_{z,\tau,-s}^{\text{KM}}(\textbf{k})$.) From Eq. (\ref{eq:Lz_KM}), we thus see that carriers with opposite spins will be associated with counter-propagating orbital currents. These (charge-neutral) orbital Hall currents are concurrent with a net spin flow; see panel (b) in Fig. \ref{supp_fig_05}. The explicit form of the OHC can be obtained by exploiting the mapping between the two models. For example, in the weak and strong scattering limits, one finds (in the metallic regime)

\begin{equation}
 \sigma_{\mathrm{oH,\mathrm{WSL}}}^{\text{s.c.}}(\varepsilon)\simeq\text{sign}(\varepsilon)\,\frac{2m_{e}v^{2}\lambda_{\textrm{KM}}^{2}}{\pi nu_{0}\varepsilon^{2}}\frac{(\varepsilon^{2}-\lambda_{\textrm{KM}}^{2})^{2}}{\left(\varepsilon^{2}+3\lambda_{\textrm{KM}}^{2}\right)^{2}},\qquad\qquad\sigma_{\mathrm{oH,\mathrm{SSL}}}^{\text{s.c.}}(\varepsilon)\simeq\frac{2m_{e}v^{2}\lambda_{\textrm{KM}}^{2}}{\pi nu_{0}\varepsilon^{2}}.
 \label{eq:OHCKM}
\end{equation}

As a second example, we consider the Haldane model of a crystalline Chern insulator. The Hamiltonian is

\begin{equation}
H_{\tau}^{(\textrm{H})}=H_{0,\tau}^{(\textrm{H})}+V^0_{\text{dis},\tau}=v\,\boldsymbol{\sigma}\cdot\mathbf{p}+\Delta_{\text{H}}\sigma_z+\sum_{i}\,u_{0}\,\sigma_{0}\,\delta(\mathbf{x}-\mathbf{x}_{i}),
\label{eq:ham_H}
\end{equation}
where  $\Delta_\text{H}=E_{\text{gap}}/2$ with $E_{\text{gap}}$ the energy gap. Note that this model breaks time-reversal symmetry since $\sigma_z\rightarrow - \sigma_z$ under $\mathcal T$ (recall that we work with the valley-isotropic representation). The reasoning developed above applies here, which allow us to quickly deduce the expressions of the various observables of interest. Specifically, the  OMM operator around a valley is

\begin{equation}
L_{z,\tau}^{\text{H}}(\mathbf{k})=-\frac{\Delta_{\text{H}}m_{e}v^{2}}{v^{2}k^{2}+\Delta_{\text{H}}^{2}}\sigma_{0}.
\label{eq:Lz_KM}    
\end{equation}

It is evident that the OHC will be given by Eq. (\ref{eq:OHCKM}) with $\lambda_{\text{KM}}$ replaced by $\Delta_{\text{H}}$. Strictly speaking, the spinless nature of the original Haldane model means that the OHC derived earlier must be halved to get the correct result,

\begin{equation}
 \sigma_{\mathrm{oH,\mathrm{WSL}}}^{\text{s.c.}}(\varepsilon)\simeq\text{sign}(\varepsilon)\,\frac{m_{e}v^{2}\Delta_{\textrm{H}}^{2}}{\pi nu_{0}\varepsilon^{2}}\frac{(\varepsilon^{2}-\Delta_{\textrm{H}}^{2})^{2}}{\left(\varepsilon^{2}+3\Delta_{\textrm{H}}^{2}\right)^{2}},\qquad\qquad\sigma_{\mathrm{oH,\mathrm{SSL}}}^{\text{s.c.}}(\varepsilon)\simeq\frac{m_{e}v^{2}\Delta_{\textrm{H}}^{2}}{\pi nu_{0}\varepsilon^{2}}.
 \label{eq:OHCH}
\end{equation}

Note that due to the broken $\mathcal T$ symmetry of the Haldane model, the transverse flow of OMM is accompanied by an anomalous Hall current. The various manifestations of the OSS mechanism are summarized in Fig. \ref{supp_fig_05}. These examples demonstrate that OSS manifests itself irrespective of the particular symmetries of the host system and requires only an OMM-active band structure in the sense explained above. The microscopic details of the disorder landscape are nonetheless crucial to determine the sign and strength of the ensuing orbital Hall currents.

\end{widetext}
\clearpage


\begin{thebibliography}{76}
\bibitem [{com({\natexlab{a}})}]{comment}%
  \BibitemOpen
  \href@noop {} {{\bibinfo {title} {The nonperturbative $T$-matrix
  level approach avoids the need for a separate evaluation of important
  diagrams absent in the standard ladder series (which is a fully perturbative
  procedure akin to the first Born approximation) like the Y-diagram. Though it
  requires a somewhat involved treatment of the renormalized vertex [Eq.
  (\ref{eq:BS_equation})], this technique is amenable to both analytical and
  numerical calculations and is applicable to generic non-interacting
  systems}}} \BibitemShut {NoStop}%
\bibitem [{com({\natexlab{b}})}]{comment2}%
  \BibitemOpen
  \href@noop {} {{\bibinfo {title} {The $J_y \propto \sigma_y$ skew-scattering term in the renormalized
  vertex [Eqs. (\ref{eq:renormalized_Jx1})-(\ref{eq:renormalized_Jx2_SSL})],
  though not required to produce a nonzero $\varrho_y$ (and hence a nonzero
  OHC), ensures that the OHC has the correct semiclassical scaling
  ($\sigma_{\text{OHE}}\propto n^{-1}$)}}\ }\BibitemShut
  {NoStop}%
\end{thebibliography}
\end{document}